\documentclass[reprint,floatfix,notitlepage,nofootinbib,twocolumn,superscriptaddress,prb]{revtex4-1}

\usepackage{amsmath}
\usepackage{amssymb}
\usepackage{graphicx}
\usepackage[tight]{subfigure}
\usepackage{enumitem}
\usepackage{soul}
\usepackage{cancel}
\usepackage{multirow}
\usepackage{tikz}
\usepackage{pifont}
\usepackage{xspace}
\usepackage{comment}

\makeatletter
\renewcommand{\p@subsection}{}
\renewcommand{\p@subsubsection}{}
\makeatother

\usepackage[english]{babel}
\makeatletter
\def\bbl@set@language#1{%
  \edef\languagename{%
    \ifnum\escapechar=\expandafter`\string#1\@empty
    \else\string#1\@empty\fi}%
  \@ifundefined{babel@language@alias@\languagename}{}{%
    \edef\languagename{\@nameuse{babel@language@alias@\languagename}}%
  }%
  \select@language{\languagename}%
  \expandafter\ifx\csname date\languagename\endcsname\relax\else
    \if@filesw
      \protected@write\@auxout{}{\string\select@language{\languagename}}%
      \bbl@for\bbl@tempa\BabelContentsFiles{%
        \addtocontents{\bbl@tempa}{\xstring\select@language{\languagename}}}%
      \bbl@usehooks{write}{}%
    \fi
  \fi}
\newcommand{\DeclareLanguageAlias}[2]{%
  \global\@namedef{babel@language@alias@#1}{#2}%
}
\makeatother
\DeclareLanguageAlias{en}{english}

\usetikzlibrary{arrows}
\usetikzlibrary{intersections}
\usetikzlibrary{shapes.geometric}
\usetikzlibrary{decorations.pathmorphing, patterns,shapes,fixedpointarithmetic}
\usetikzlibrary{decorations.markings}

\tikzset{
  mid arrow/.style={postaction={decorate,decoration={
        markings,
        mark=at position .575 with {\arrow{stealth}}
      }}},
  end arrow/.style={postaction={decorate,decoration={
        markings,
        mark=at position 1 with {\arrow{stealth}}
      }}},
  snake arrow/.style={fixed point arithmetic, decorate, decoration={snake,amplitude=2pt, segment length=11pt},postaction={decoration={markings,mark=at position 0.625 with {\arrow{stealth}}},decorate}},
}

\newcommand{\red}[0]{\color{red}}

\usepackage{bm}

\usepackage[pdfusetitle,colorlinks=true,citecolor=blue,linkcolor=magenta]{hyperref}

\graphicspath{{./}{./images/}}

\begin{document}

\title{Entanglement structure in the volume-law phase of hybrid quantum automaton circuits}
\author{Yiqiu Han}
\email{hankq@bc.edu}
\affiliation{Department of Physics, Boston College, Chestnut Hill, MA 02467, USA}

\author{Xiao Chen}
\email{chenaad@bc.edu}
\affiliation{Department of Physics, Boston College, Chestnut Hill, MA 02467, USA}

\begin{abstract}
 We study entanglement fluctuations and quantum error correction in the weakly monitored volume-law phase of quantum automaton circuits subject to repeated local measurements. We numerically observe that the entanglement entropy exhibits strong fluctuation with the exponent close to the ``growth exponent'' of the Kardar-Parisi-Zhang (KPZ) universality class, the same as other local random circuits studied previously. We also investigate the dynamically generated quantum error correction code in the purification process and show that this model has different contiguous code distances for two types of errors that exhibit similar sublinear power-law scaling. We give an interpretation of these results by mapping them to various quantities in a classical particle model. We demonstrate that the subleading correction term of the entanglement entropy and the sublinear power-law scaling of the contiguous code distance in the volume-law phase are both the emergent phenomena of the hybrid random dynamics. Finally, we show that this classical particle dynamics itself has a type of error correction ability and can dynamically generate a classical linear code.

\end{abstract}

\maketitle
\section{Introduction}
    The past few years have witnessed a surge of interest in monitored quantum dynamics\cite{PhysRevX.9.031009,PhysRevB.99.224307,PhysRevB.98.205136,PhysRevB.100.134306,PhysRevX.10.041020,PhysRevB.101.104301,PhysRevB.101.104302,Chen_2020,Ippoliti_2021,Sang_2021,Lavasani_2021}. These nonunitary dynamics can exhibit many emergent phenomena which are inaccessible in pure unitary dynamics or in systems in equilibrium. It is by now well-known that in a generic interacting system, repeated measurements can induce a continuous phase transition from a highly entangled volume-law phase to a disentangled area-law phase \cite{PhysRevX.9.031009,PhysRevB.99.224307,PhysRevB.98.205136,PhysRevB.100.134306,PhysRevX.10.041020,PhysRevB.101.104301,PhysRevB.101.104302}. In addition, specific types of measurements can stabilize various quantum phases, including critical phases and ordered phases \cite{Chen_2020,PhysRevB.105.064306,Ippoliti_2021,Sang_2021,Lavasani_2021}. These rapid developments significantly broaden our understanding of nonequilibrium dynamics. 

    To understand these emergent phenomena in monitored quantum dynamics, various nonunitary random circuits have been constructed. This includes hybrid random Clifford circuits and hybrid random Haar circuits. For Clifford circuits, there exists a very efficient algorithm in terms of the stabilizer formalism which allows us to simulate nonunitary dynamics for very large system sizes \cite{GottesmanKnill,PhysRevB.98.205136,PhysRevB.100.134306,PhysRevX.10.041020}. On the other hand, Haar circuits provide an important analytical approach which can map many quantum dynamics problems to statistical mechanics models \cite{PhysRevX.9.031009,PhysRevB.101.104301,PhysRevB.101.104302}. 
    
    Recently, a new type of circuit called hybrid quantum automaton (QA) circuit was constructed to investigate the entanglement dynamics in the monitored quantum systems \cite{Iaconis_2021pmf}. This circuit is composed of QA unitaries and local composite measurements. The detail of these two types of gates will be explained later in the paper. Compared with random Haar/Clifford circuits, QA circuits not only provide an efficient method for large-scale numerical simulation, but also provide an analytical tool to understand the quantum dynamics. 
    Due to the basis-preserving feature of QA circuits, the entanglement dynamics can be interpreted in terms of a classical bit-string picture. Specifically, the second R\'enyi entropy can be mapped to the first-passage problem in the bit-string dynamics. Based on this mapping, it was further shown that the measurement-induced entanglement phase transition in a generic hybrid QA circuit belongs to the directed percolation (DP) universality class\cite{Iaconis_2021pmf}. At the critical point, the prefactor of the logarithmic entanglement is related to the local persistent exponent in the DP universality class. By further imposing symmetries in the dynamics, new critical points or critical phases belonging to different universality classes can also be identified \cite{PhysRevB.105.064306}. 
    
    Aside from the entanglement phase transition at the critical point, the volume-law phase itself also has an interesting entanglement structure. Previous studies for one-dimensional (1d) random Haar circuits suggested that the entanglement entropy can be mapped to the free energy of the directed polymer in a random environment (DPRE) which has fluctuation belonging to the Kardar-Parisi-Zhang (KPZ) universality class \cite{PhysRevX.7.031016,PhysRevX.9.031009,KPZ,PhysRevLett.58.2087}. Such fluctuations lead to a subleading correction term scaling as $L^{1/3}$ in the entanglement entropy in the volume-law phase, with $L$ being the system size. This has also been numerically verified for 1d random Clifford circuits \cite{li2021entanglement,PhysRevB.103.104306}. Interestingly, it is found in 1d random Haar circuits coupled with dephasing channels on the boundary that the entanglement negativity in the steady state has a \textit{leading} $L^{1/3}$ power-law scaling for $0<p<p_c$ \cite{EhudNegativity}. Inspired by the above works, in this paper we will study the entanglement properties of the weakly monitored volume-law phase of 1d hybrid QA circuits. In particular, we study the fluctuation of the entanglement entropy and the quantum error correction property of the volume-law phase. We numerically show that the entanglement entropy of a subsystem in both the early time dynamics and the steady state in the hybrid Clifford QA circuit also exhibit sample-to-sample fluctuations with the same scaling behavior. In order to understand this behavior, we construct a classical two-species particle model based on the bit-string picture in which the particles undergo stochastic random dynamics. However, the classical quantity in the two-species particle model which corresponds to the second R\'enyi entropy in the QA circuit is only numerically simulable for the early time dynamics which exhibits $t^{1/3}$ fluctuation in the time direction in the weakly monitored phase. To solve this issue, we propose two approximations, namely, the single-species and the approximated two-species particle model, to estimate the fluctuations more efficiently.

    
    In addition, we study the purification dynamics of a maximally-mixed initial state in the volume-law phase of hybrid QA circuits \cite{PhysRevX.10.041020}. We modify the aforementioned particle model slightly and use this to give an interpretation of the entanglement entropy of a subsystem in the presence of the environment. Previously, it was shown that for purification dynamics, the hybrid quantum circuit can dynamically generate a quantum error correcting code (QECC) \cite{PhysRevX.10.041020,SoonwonQECC,RuihuaQECC}. The contiguous code distance of the QECC, defined as the minimum length of a contiguous subsystem that supports an uncorrectable error, was quantitatively investigated in the random Clifford circuit and found to exhibit a power-law scaling $L^{\gamma}$ with $\gamma\approx\frac{1}{3}$ \cite{PhysRevB.103.104306}. We analyze the quantum error-correcting property of the hybrid QA circuit and explain it in terms of the particle model. In particular, it has two types of contiguous code distance for different errors occurring in the quantum system. We show that both of them have a similar power-law scaling with the exponent close to $\frac{1}{3}$. In addition, we show that both the sublinear power-law exponent in the code distance and the fluctuation exponent in the entanglement entropy are the results of hybrid random dynamics.

    
    Interestingly, the stochastic classical particle model itself has an error correction property, and can dynamically generate a classical linear code (CLC). We study this random CLC by analyzing the dynamics of the associated generator matrix and numerically compute its contiguous code distance. 
    
\section{Review of the hybrid QA circuit and two-species particle model}

Ref.~\onlinecite{Iaconis_2021pmf} establishes the relationship between the entanglement dynamics and the classical bit-string dynamics in the hybrid QA circuit. The subsequent work of Ref.~\onlinecite{PhysRevB.105.064306} explicitly constructs a classical two-species particle model to describe the entanglement dynamics of the $\mathbb{Z}_2$-symmetric hybrid QA circuit. In this section, we briefly review some of the important results in these two papers and modify the two-species particle model so that it can be applied on hybrid QA circuits without any symmetry.

The hybrid QA circuit is composed of QA unitary operators and composite measurements. A QA unitary gate permutes product states in the computational basis up to a phase, i.e.,
\begin{equation}
  U|n\rangle=e^{i\theta_n}|\pi(n)\rangle,
\end{equation}
where $\pi\in S_{2^L}$ is an element of the permutation group on the computational basis of a lattice with $L$ qubits. We choose the Pauli Z basis as the computational basis and take the initial state to be a product state polarized in $x$ direction, $|\psi_0\rangle=|+x\rangle^{\otimes L}$. 
QA unitaries scramble the phase information stored in the wave function and hence increase the entanglement of the state until it saturates to the volume-law scaling. Meanwhile, the wave function remains an equal-weight superposition of computational basis states, which is the characteristic of QA circuits.

On the other hand, local measurements can suppress the growth of entanglement. In the QA circuit, the composite measurement of the $i$-th qubit is defined as a projection operator followed by a Hadamard gate,
\begin{equation}
  M_i^\sigma=H_i\circ P_i^\sigma,
  \label{eq:cm}
\end{equation}
where $P_i^\sigma=\frac{1\pm Z_i}{2}$ is the Pauli Z measurement on site $i$ with the outcome $\sigma=\{0,1\}$ and $H_i$ rotates the state back to an equal-weight superposition over the computational basis. Therefore, after imposing $M_i^\sigma$, the phase information for half of the basis states is lost. The composite measurements disentangle the system while preserving the special feature of QA circuits.

It is shown in Ref.~\onlinecite{Iaconis_2021pmf} that by increasing the measurement rate $p$, the one-dimensional hybrid QA circuit undergoes an entanglement phase transition from a volume-law entangled phase to an area-law disentangled phase, with the phase transition belonging to the $1+1$d directed percolation (DP) universality class. If we bipartition the system into subsystem $A$ and its complement $B$, a common quantity to measure the entanglement between them is the $n^{th}$ R\'enyi entropy:
\begin{equation}
  \begin{aligned}
    S^{(n)}_A &=\frac{1}{1-n}\log_2{[\text{Tr}(\rho_A^n)]}
    \\
    \rho_A &=\text{Tr}_{B} |\psi\rangle\langle\psi|.
  \end{aligned}
\end{equation}
In this paper, we focus on the second R\'enyi entropy with $n=2$, whose purity equals the expectation value of the $\mathsf{SWAP}_A$ operator over two copies of the state \cite{PhysRevLett.104.157201,islam2015measuring},
\begin{equation}\label{eq: swap}
  \text{Tr}[\rho_A^2(t)]=\langle\psi_t|_2\otimes\langle\psi_t|_1 \mathsf{SWAP}_A|\psi_t\rangle_1\otimes|\psi_t\rangle_2,
\end{equation}
 with the wave function $|\psi_t\rangle=\tilde{U}_t|\psi_0\rangle$, where $\tilde{U}_t$ denotes the circuit evolution up until time $t$ (For more details, see Appendix~\ref{Appendix: 2ps}). The $\mathsf{SWAP}_A$ operator exchanges the spin configurations within subsystem $A$ of $|\psi_t\rangle_1$ and $|\psi_t\rangle_2$.

The entanglement dynamics of the hybrid QA circuit can be interpreted in terms of classical bit-string dynamics. By inserting two sets of complete basis which we call ``bit strings'' on the right side of the $\mathsf{SWAP}_A$ operator in Eq. \ref{eq: swap} and applying the circuit on the bit-strings in a time-reversed order, we obtain
\begin{equation}
  \begin{aligned}
    \text{Tr}[\rho_A^2(t)]&=\sum_{n_1,n_2}\langle\psi_t|_2\langle\psi_t|_1 \mathsf{SWAP}_A|n_1\rangle|n_2\rangle\langle n_2|\langle n_1|\psi_t\rangle_1|\psi_t\rangle_2
    \\
    &=\frac{1}{4^L} \sum_{n_1,n_2}e^{-i\Theta_{n_1'}(t)}e^{-i\Theta_{n_2'}(t)}e^{i\Theta_{n_1}(t)}e^{i\Theta_{n_2}(t)},
  \end{aligned}
    \label{eq:N_A}
\end{equation}
where 
\begin{equation}
    e^{i\Theta_{n_i}(t)}=\sqrt{2^L}\langle n_i|\tilde{U}_t|\psi_0\rangle,
\end{equation}
and
\begin{equation}
  \begin{aligned}
    |n_1'\rangle|n_2'\rangle&\equiv \mathsf{SWAP}_A|n_1\rangle|n_2\rangle
    \\
    &=\mathsf{SWAP}_A|\alpha_1\beta_1\rangle|\alpha_2\beta_2\rangle
    \\
    &=|\alpha_2\beta_1\rangle|\alpha_1\beta_2\rangle,
  \end{aligned}
\end{equation}
where $|\alpha_i\rangle$ and $|\beta_i\rangle$ are the spin configurations in subsystems $A$ and $B$ of $|n_i\rangle$. In order to compute $\mbox{Tr}(\rho_A^2)$, we need to understand the dynamics of the relative phase $\Theta_r=-\Theta_{n_1'}-\Theta_{n_2'}+\Theta_{n_1}+\Theta_{n_2}$ for each bit-string pair $\{|n_1\rangle,|n_2\rangle\}$. Under QA evolution, nonzero randomly distributed $\Theta_r$ will lead to destructive interference, and as such only configurations with trivial relative phase contribute to the purity. This observation motivates us to construct a two-species particle model \cite{PhysRevB.105.064306}. 

\begin{figure}
  \centering
  \includegraphics[width=0.3\textwidth]{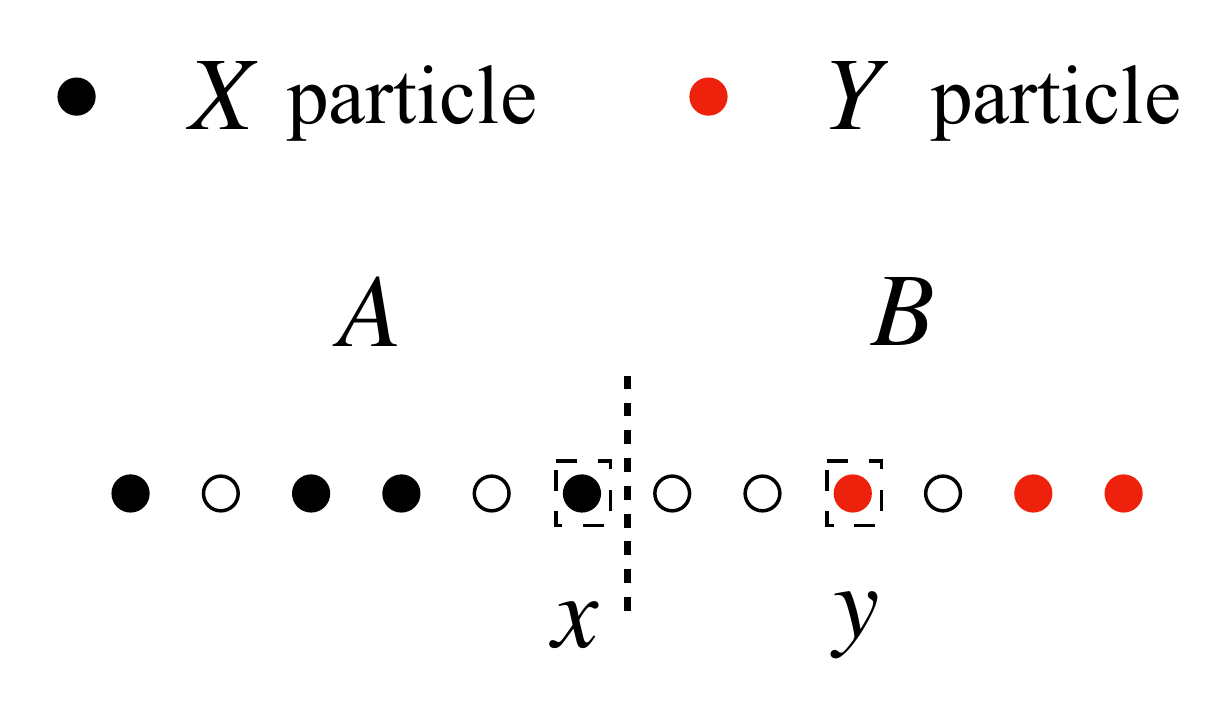}
  \caption{A cartoon of the two-species particle model. The black dots represent $X$ particles, and the red dots represent $Y$ particles. Initially, $X$ and $Y$ particles are distributed in region $A$ and $B$ respectively. We use $x$ and $y$ to denote the leftmost $X$ particle and the rightmost $Y$ particle.}
  \label{fig:2ps_cartoon1}
\end{figure}

The particles here characterize the difference between a bit-string pair $\{|n_1\rangle,|n_2\rangle\}$,
\begin{equation}
  h(x,t)=|n_1(x,t)-n_2(x,t)|.
\end{equation}
We use the empty site symbol $\circ$ to denote $h(x)=0$ and the occupied site symbol $\bullet$ to denote $h(x)=1$. Specifically, we represent the difference at $t=0$ in $A$ ($B$) by $X$ ($Y$) particles, as illustrated in Fig. \ref{fig:2ps_cartoon1}. Under the time evolution, these two species start to expand according to the update rule determined by the circuit. For the rest of the paper, we focus on QA unitary gates $U$ which are linear with respect to the bit-string addition defined in finite field $\mathbb{F}_2$, so that for any bit string pair $n_1$ and $n_2$, $U(|n_1\rangle+|n_2\rangle)=U|n_1+n_2\rangle$. This means that we can directly work on the particle representation $h(x,t)$ without keeping track of the bit-string dynamics. One good example is the two-qubit CNOT gate. When the first qubit acts as the control, we have $\bullet\circ\leftrightarrow\bullet\bullet$. On the other hand, the composite measurement forces the spins on the same site to be identical, which results in particle annihilation, $\bullet\to\circ$. As shown in Appendix \ref{Appendix: 2ps}, only the configurations in which the $X$ and $Y$ particles do not meet up to time $t$ yield $\Theta_r(t)=0$ and hence contribute to the purity. Therefore, we have
\begin{equation}
  \begin{aligned}
    &\text{Tr}\rho_A^2(t)= \frac{N(t)}{2^L}\equiv P(t),
    \\
    &S_A^{(2)}(t)= -\log_2 P(t),
  \end{aligned}
  \label{eq:dyn}
\end{equation}
where $N(t)$ is the number of configurations in which the two species do not meet up to time $t$. At the critical point $p=p_c$, the fraction $P(t)$ decays algebraically as $P(t)\propto t^{-\alpha}$, where $\alpha$ is the persistence exponent. For the DP universality class, $\alpha=0.938$ is a universal number \cite{Iaconis_2021pmf}. This power-law decay is responsible for the logarithmic scaling of the entanglement entropy at criticality.

\section{Entanglement dynamics in the volume-law phase}

We now take a closer look at the entanglement entropy in the volume-law phase with $p<p_c$. Previous studies of these $1+1$d hybrid circuits indicate that randomness induces strong fluctuations in the entanglement entropy in both spatial and temporal directions.
A nice way to understand this problem is through the minimal cut picture introduced in Ref.~\onlinecite{PhysRevX.7.031016}, which maps the entanglement dynamics to the first passage problem on a percolation lattice. Such a picture rigorously describes the zeroth R\'enyi entropy $S_A^{(0)}$ of the Haar random circuit subject to projective measurements. For the entropy with higher R\'enyi index, it is argued that it can be treated as the free energy of the domain wall in a disordered magnet\cite{li2021entanglement,Zhou_2019}. Notice that in both approaches, the entanglement entropy is mapped to the free energy of the $1+1$d directed polymer in a random environment (DPRE), whose fluctuation belongs to the KPZ universality class.  As a result, there exists a sub-leading correction term in the ensemble averaged entanglement entropy in both the early time dynamics and the steady states, i.e.,
\begin{align}
    &\langle S_A(t)\rangle =\lambda_1 t+ a t^{\beta}+\cdots,
    \\
    &  \langle S_A(L_A)\rangle=\lambda_2 L_A+ b L_A^{\beta}+\cdots,
\end{align}
where the brackets represent an ensemble average and $\beta=\frac{1}{3}$ is the ``roughness exponent'' of the DPRE\cite{KPZ}. The sub-leading correction term can be extracted by computing the standard deviation
\begin{align}
&\delta S_A(t)=\sqrt{\langle[S_A(t)]^2\rangle-\langle S_A(t)\rangle^2} \propto t^{\beta},\\
 & \delta S_A(L_A)=\sqrt{\langle[S_A(L_A)]^2\rangle-\langle S_A(L_A)\rangle^2} \propto L_A^{\beta},
\end{align}
which characterizes the sample-to-sample fluctuations with the same exponent $\beta$. This result has been confirmed numerically in Clifford circuits in Refs.~\onlinecite{li2021entanglement,PhysRevB.103.104306}. Below we will numerically examine the volume-law phase of the hybrid Clifford QA circuit and understand its physics in terms of the particle dynamics. 

\subsection{Numerical study in hybrid Clifford QA circuits}
\begin{figure}[tp!]
  \centering
  \subfigure[]{
    \includegraphics[width=.3\textwidth]{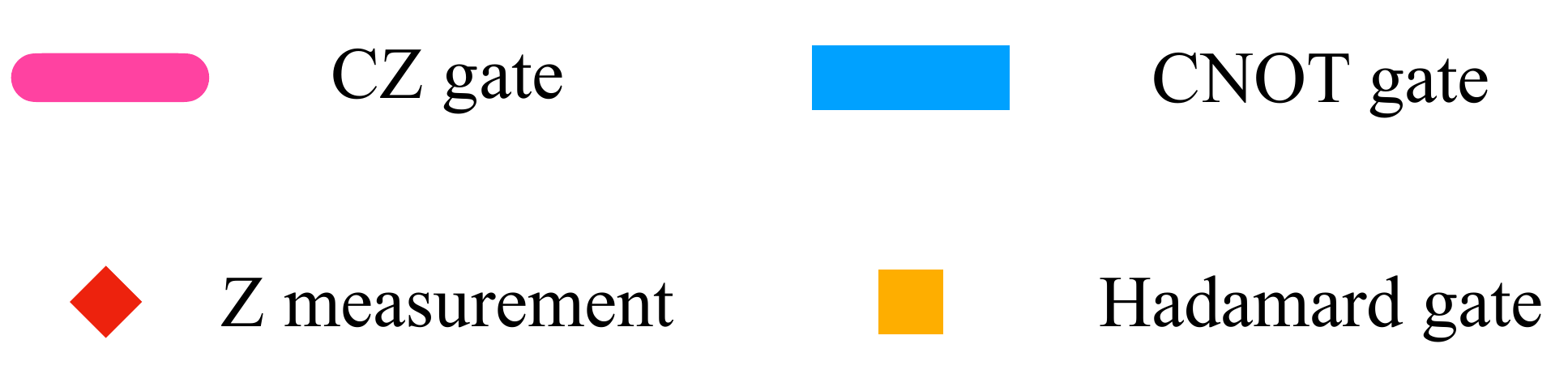}
    \label{fig:EE_gate}}
  \subfigure[]{
    \includegraphics[width=.3\textwidth]{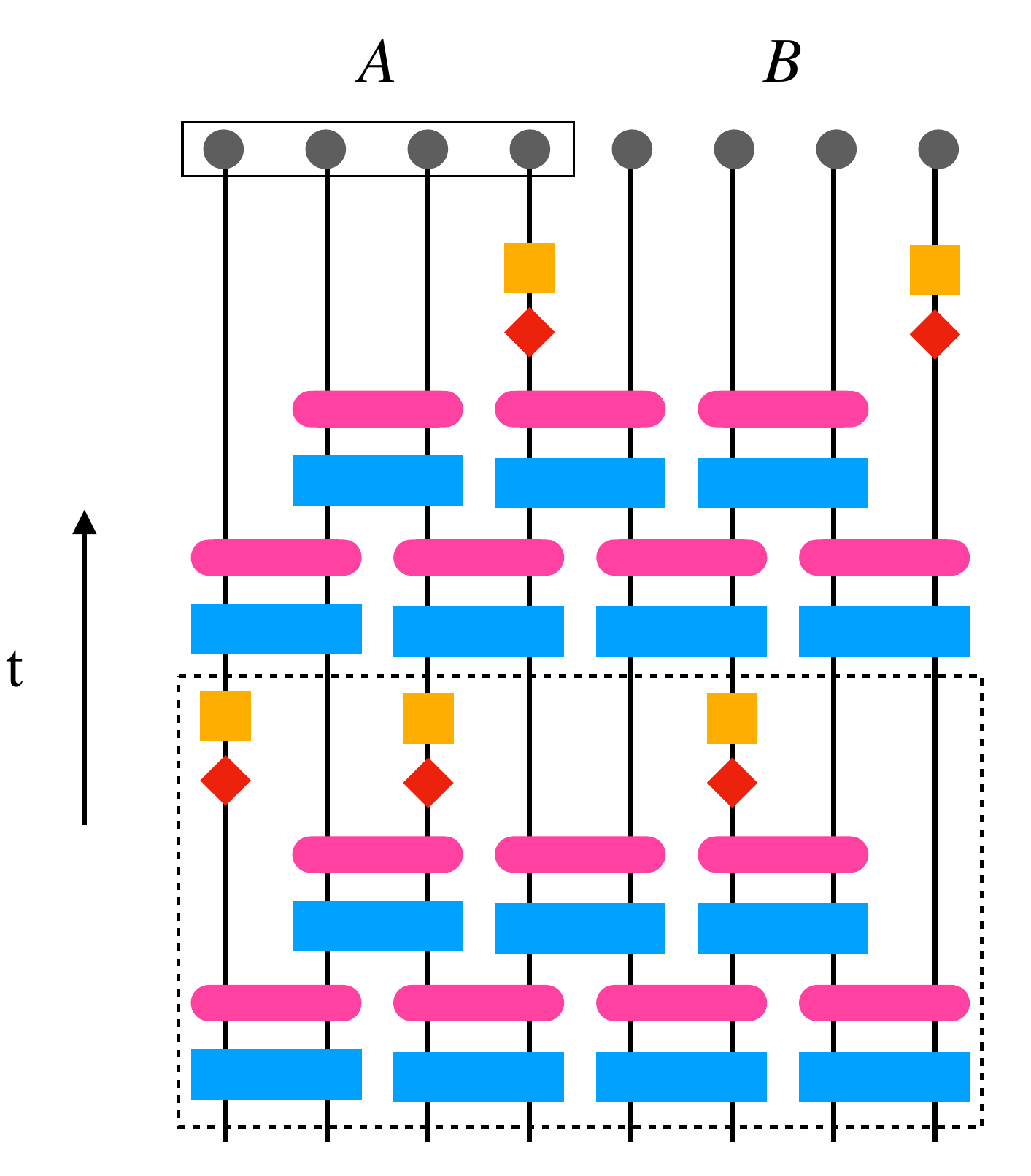}
    \label{fig:QA_EE}}

  \caption{(a) A schematic for the gates appearing in the hybrid Clifford QA circuit. (b) The dashed box represents the arrangement of gates in a single time step. Each time step involves two layers of CNOT gates and two layers of CZ gates, interspersed with composite measurements with probability $p$.}
  \label{fig:EE_circuit}
\end{figure}

\begin{figure}[t!]
  \centering
  \subfigure[]{
    \includegraphics[width=.4\textwidth]{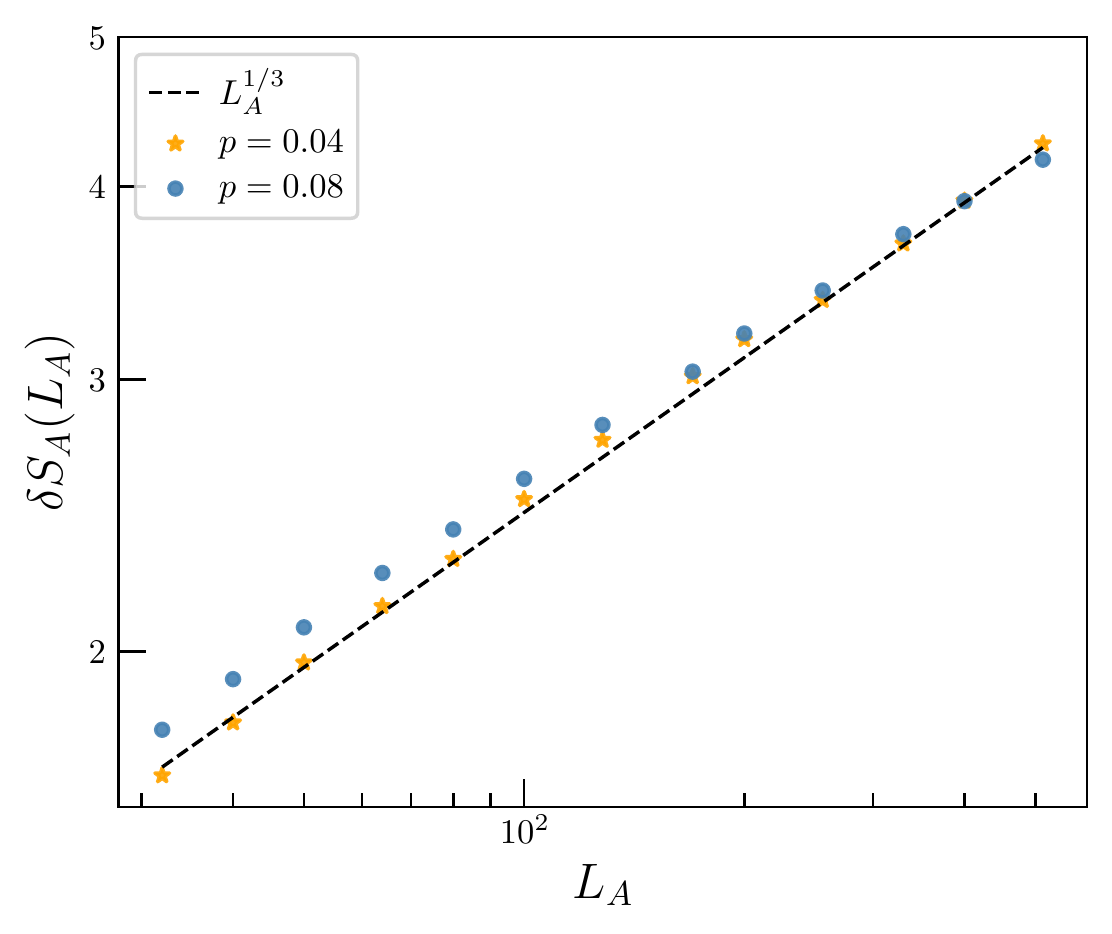}
    \label{fig:EE_LA_delta_SA}}
  \subfigure[]{
    \includegraphics[width=.4\textwidth]{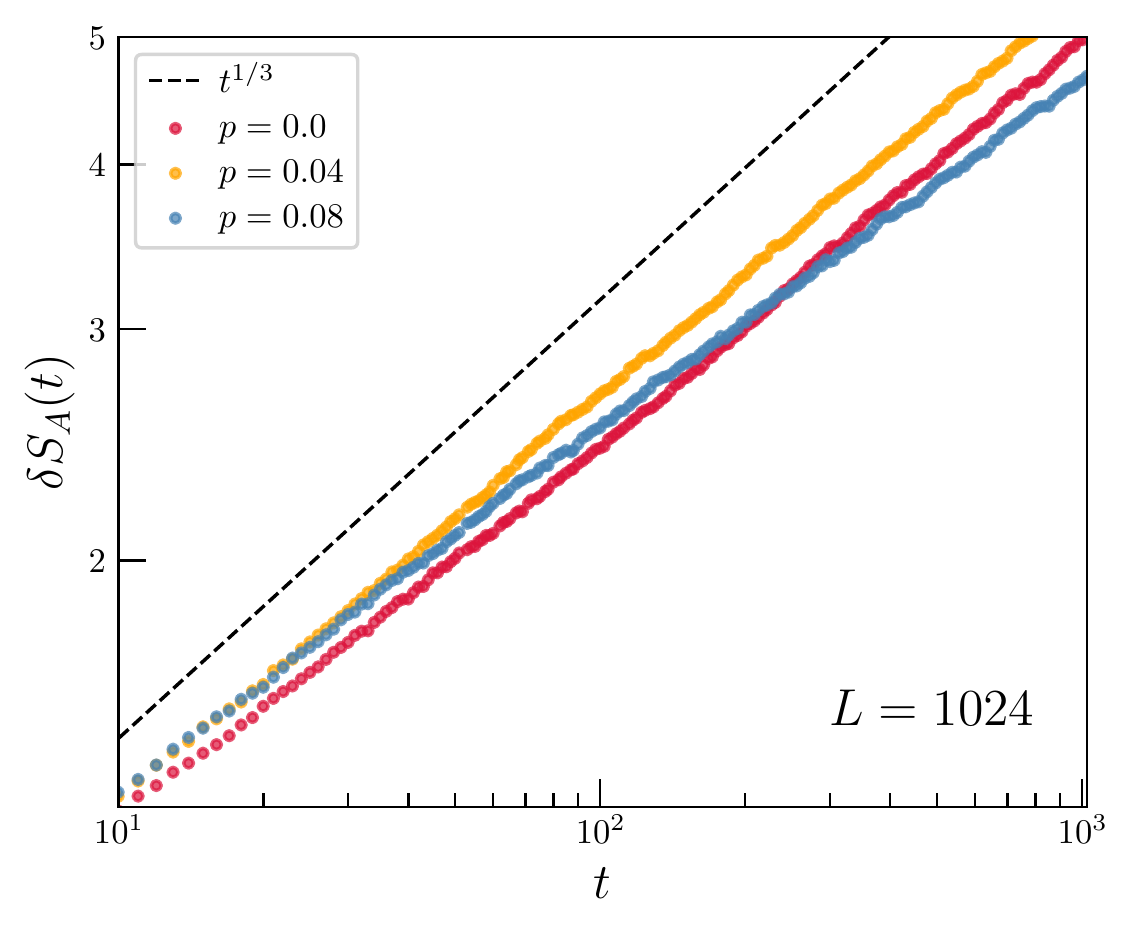}
    \label{fig:EE_t_delta_SA}}

  \caption{(a) The standard deviation of entanglement entropy $\delta S_A$ vs $L_A$ plotted on a log-log scale. The data are computed from the steady-state entanglement entropy $S_A$ for half-system size $L_A=L/2$ over a variety of $L$. We find that $\delta S_A\propto L_A^{\beta_1}$ with $\beta_1\approx 0.322$ at $p=0.04$ and $\beta_2\approx 0.31$ at $p=0.08$. (b) The standard deviation of early time entanglement entropy $\delta S_A$ vs $t$ for $p=0,0.04,0.08$, where we find $\delta S_A(t)\propto t^{\beta_2}$ with $\beta_2\approx 0.307$ for $p=0$ and $p=0.04$, and $\beta_2\approx 0.266$ for $p=0.08$. All of the numerical data for entanglement entropy are calculated with periodic boundary conditions (PBC).}
\end{figure}

We consider a hybrid Clifford QA circuit in which the QA unitaries also belong to the Clifford group. According to the Gottesman-Knill theorem \cite{GottesmanKnill,PhysRevA.70.052328}, the Clifford circuit can be efficiently simulated on a classical computer using the stabilizer formalism. As illustrated in Fig. \ref{fig:EE_circuit}, the circuit is constructed from two types of unitary gates chosen from the two-qubit Clifford group, namely, CNOT and CZ gates, as well as sporadic composite measurements distributed with probability $p$. The critical point is at $p_c\approx 0.138$ \cite{Iaconis_2021pmf}. In the numerical simulation, we take $p=0.04$ and $p=0.08$ to investigate the  fluctuation of volume-law phase entanglement entropy. The numerical results are averaged over $O(10^4)$ samples. Despite the uncertainty caused by the data itself, there still exists an uncertainty when estimating the fitted exponents of $\delta S_A$, which depending on the number of data points, can have an error up to $\pm 0.04$. Therefore in the figures where the scaling exponents are close to $\beta=\frac{1}{3}$, we will only plot the curve $L_A^{1/3}$ or $t^{1/3}$ with a constant offset for comparison and the estimated exponents are given in the text and summarized in Table. \ref{table:exponents}. Our numerical results in Fig. \ref{fig:EE_LA_delta_SA} indicate that for $p=0.04$ and $p=0.08$, the standard deviation of the steady-state entanglement entropy scales as $\delta S_A\propto L_A^{\beta_1}$ with $\beta_1\approx 0.322$ and $0.31$ respectively.

Aside from the steady state, we also study the early time entanglement dynamics in the volume-law phase. Similarly, we observe in Fig. \ref{fig:EE_t_delta_SA} that for $p<p_c$ (not necessarily nonzero), $\delta S_A(t)\propto t^{\beta_2}$ with $\beta_2\approx 0.307$ for $p=0$ and $p=0.04$, and $\beta_2\approx 0.266$ for $p=0.08$. For $p\leq 0.04$, The exponents of the sub-leading terms of the steady-state and early time entanglement entropies are similar and are close to the roughness exponent, i.e., $\beta_1\approx \beta_2\approx \frac{1}{3}$. The exponent $\beta_2\approx 0.266$ at $p=0.08$ is smaller than $1/3$ and is probability due to the proximity to the critical point. We also consider the entanglement entropy in the purification dynamics of the same circuit and in the $\mathbb{Z}_2$-symmetric hybrid Clifford QA circuit. In both cases, we find that it exhibits similar fluctuation. The details can be found in Appendixes ~\ref{Appendix: Clifford puri} and ~\ref{Appendix: Z_2}. Overall, these results provide numerical evidence that the entanglement entropy in the volume-law phase of the Clifford QA circuit has KPZ fluctuations.

\subsection{Single-species particle dynamics}\label{sec: 1ps}

\begin{figure}[tp!]
  \centering
  \subfigure[]{
    \includegraphics[width=0.4\textwidth]{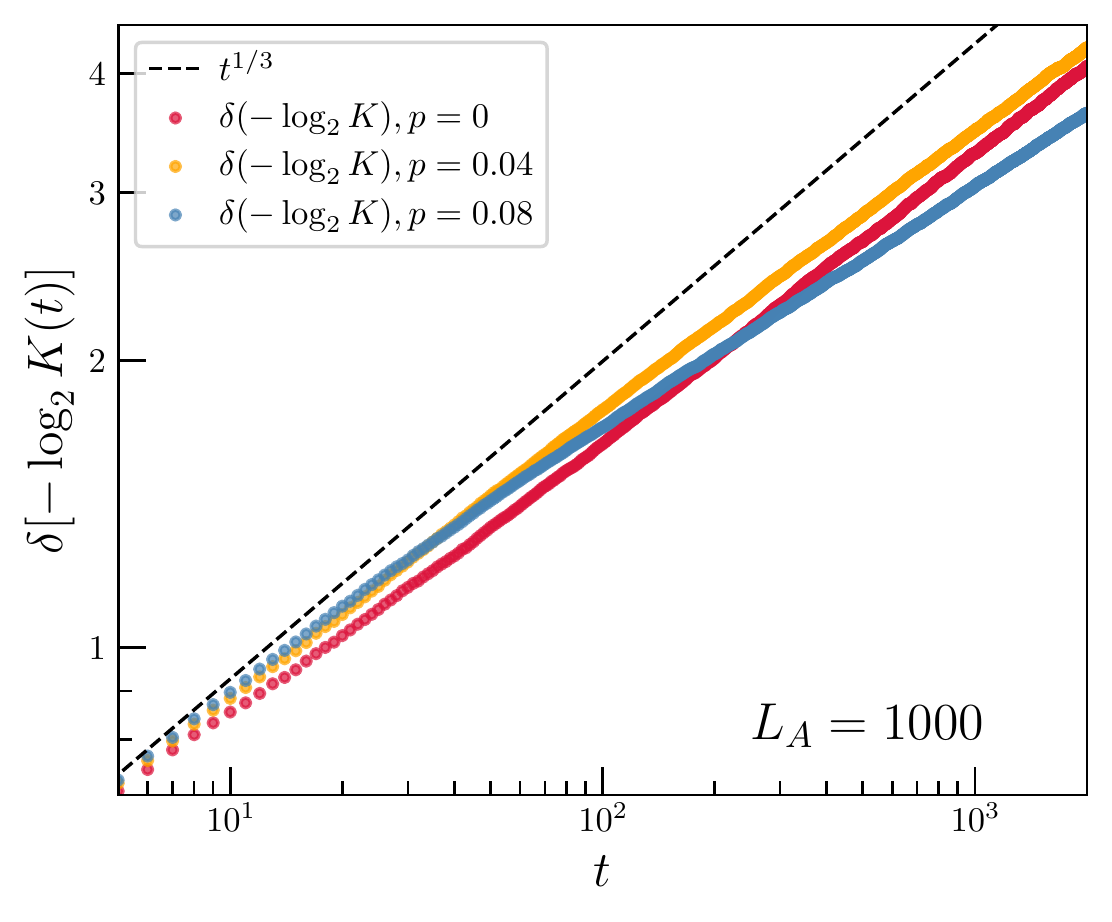}
    \label{fig:1ps_t}}
  \subfigure[]{
    \includegraphics[width=0.4\textwidth]{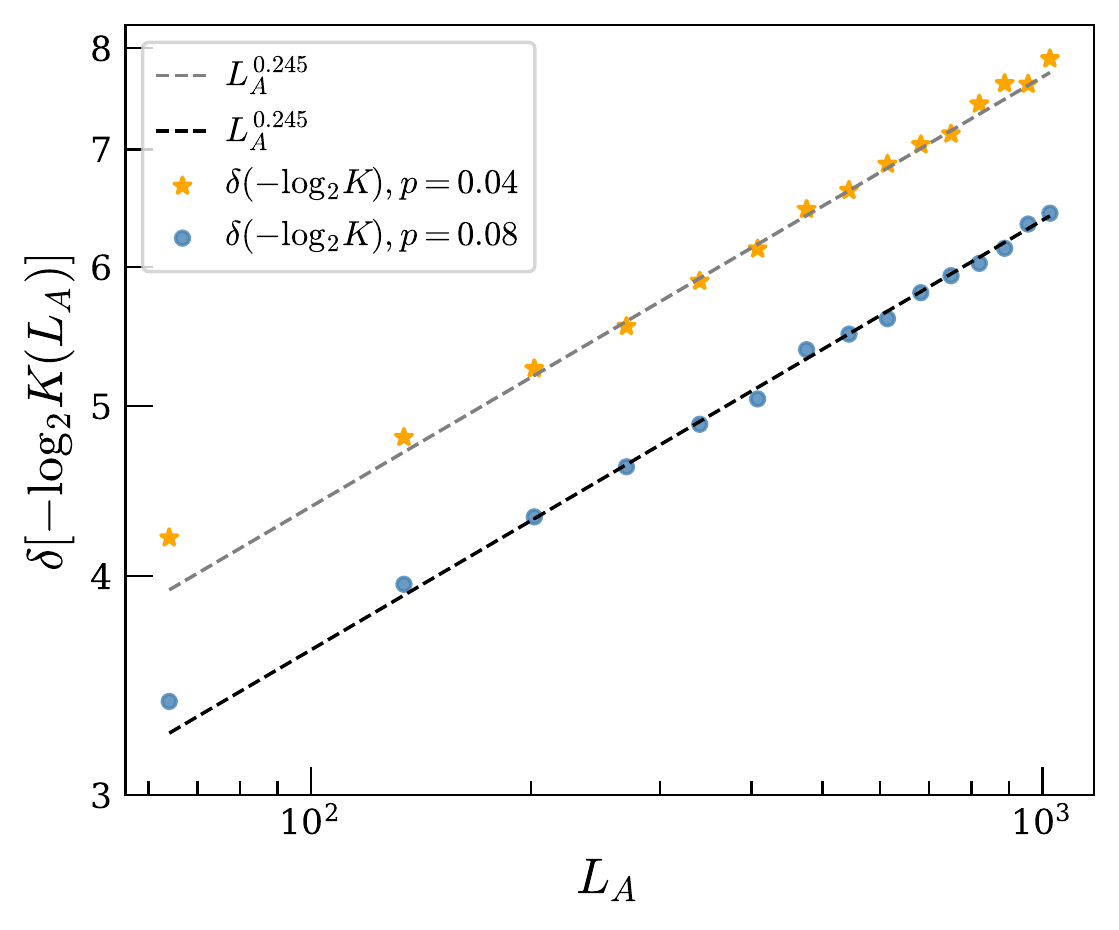}
    \label{fig:1ps_LA}}
  \caption{(a) The early time $\delta(-\log_2{K})$ vs $t$ plotted on a log-log scale. We find that $\delta[-\log_2{K(t)}]\propto t^{0.304}$ at $p=0$, $t^{0.294}$ at $p=0.04$, and $t^{0.26}$ at $p=0.08$. (b) The steady state $\delta(-\log_2{K})$ of the single-species particle model vs $L_A$ plotted on a log-log scale. The numerical data are calculated from the single-species particle model using the basis-decomposing method with particle annihilation rate $p=0.04$ and $p=0.08$.}
\end{figure}

Recall that in the two-species particle model, the entanglement entropy is related to the logarithm of $P(t)$, which is the fraction of configurations in which $X$ particles do not encounter $Y$ particles up to $t$. We denote $x$ ($y$) as the rightmost $X$ (leftmost $Y$) particle. In the volume-law phase, $x$ and $y$ move toward each other at roughly the same speed, so $P(t)$ decays exponentially in $t$, leading to the linear growth of $S_A^{(2)}(t)$. The subleading term in $S_A^{(2)}(t)$ is caused by the fluctuation of the velocities of $x$ and $y$ in different particle configurations. For simplicity, we fix the position of $y$ to be next to the boundary between $A$ and $B$, so that only the fluctuation of $x$ is considered. This is equivalent as computing a subset of phase terms in Eq.\eqref{eq:N_A} restricted in subsystem $B$, 
\begin{align}
    \frac{1}{4^{L_A}}\sum_{\alpha_1,\alpha_2}e^{-i\Theta_{n_1^\prime}^{B}+i\Theta_{n_1}^{B}}.
\end{align}
where $\{\alpha_{1},\alpha_2\}$ are the spin configurations in subsystem $A$ of the bit-string pairs $\{|n_1\rangle,|n_1^\prime\rangle\}$. With this approximation, $P(t)$ is simplified to be $K(t)$, the fraction of configurations in which $x$ never crosses the boundary between $A$ and $B$ up to time $t$.

One important advantage of taking the single-species approximation is that $K(t)$ can be efficiently computed using the following approach. (1) All of the particle configurations in subsystem $A$ can be generated by a set of binary basis $H^0=\{h_1,\dots,h_{L_A}\}$. Hence, any particle configuration can be expressed as the linear combination
\begin{align}
h=\sum_{i=1}^{L_A}h_i^{\alpha_i}
\end{align}
defined on the finite field $\mathbb{F}_2$ with $\alpha_i=\{0,1\}$. 
Initially, we can set $h_i(t=0)=(0\dots1_i\dots0)$. Under linear operators, we can evolve each basis separately and the above equation still holds with $\{\alpha_i\}$ remaining invariant. (2) $K(t)$ can be evaluated by simply evolving a set of basis $H(t)$ which span the ensemble of particle configurations which never enter $B$. Initially, $H(t=0)=H^0$ and therefore $K(t=0)=1$. Under the time evolution, if the rightmost particle $x$ of a single basis state, say, $h_j(t)$, crosses the boundary, then only the particle configurations with $\alpha_j=0$ will contribute to $K(t)$. Hence, half of the configurations are ruled out,  and the ``entanglement entropy'' $-\log_2{K(t)}$ increases by one. This also means that $h_j$ is excluded from $H(t)$ for further computation. On the other hand, if the $x$ particles of multiple basis states, say, $G=\{h_1,\dots,h_n\}$, cross the boundary at the same time, one can easily verify that $-\log_2{K(t)}$ still increases by one, except that the updated basis set becomes $H(t)=\{h_1+h_2,\dots,h_1+h_n\}\bigcup H(t-1)\setminus G$. As a result, \begin{align}
-\log_2{K(t)}=L_A-|H(t)|,
\end{align}
where $|H(t)|$ is the number of existing basis at time $t$. This way of tracing the basis set which span the configurations whose boundary has not been visited by the particles resembles the stabilizer formalism in Clifford dynamics.

We use the above method to first study the $p=0$ limit of the single-species particle model under the Clifford QA circuit defined in Fig. \ref{fig:QA_EE}. With this limit, the particle basis states evolve under only unitary operators, i.e., random CNOT gates. The numerics in Fig. \ref{fig:1ps_t} shows that the early time dynamics has the fluctuation $\delta[-\log_2 K(t)]\propto t^{0.304}$. In the steady state, the particles in all the basis states will pass the boundary and therefore $-\log_2 K(t\to\infty)=L_A$ without subleading correction. 

When $0<p<p_c$, we observe similar fluctuations in the early time dynamics. Fig. \ref{fig:1ps_t} shows that $\delta[-\log_2{K(t)}]\propto t^{0.294}$ at $p=0.04$ and $\delta[-\log_2{K(t)}]\propto t^{0.26}$ at $p=0.08$. The power law exponent slightly decreases as we increase $p$. As opposed to the $p=0$ case, the steady state entropy cannot reach the maximal value. Due to the local measurement which forces $\bullet\to\circ$ in all the basis states at the same location, the time-evolved basis states in $H(t)$ cease to remain mutually linearly independent. The steady state basis vectors $H(t\to\infty)$ are formed by zero vectors containing no particles. The fluctuation of the number of such zero vectors is the same as the fluctuation of $-\log_2 K(t\to\infty)$ and is shown in Fig. \ref{fig:1ps_LA}. By performing finite size scaling, we observe that $\delta[-\log_2{K(L_A)}]\propto L_A^{0.245}$ for both $p=0.04$ and $p=0.08$, slightly off from $1/3$.

\subsection{Two-species particle dynamics}\label{sec: 2ps}
\begin{figure}[tp!]
  \centering
  \subfigure[]{
    \includegraphics[width=0.4\textwidth]{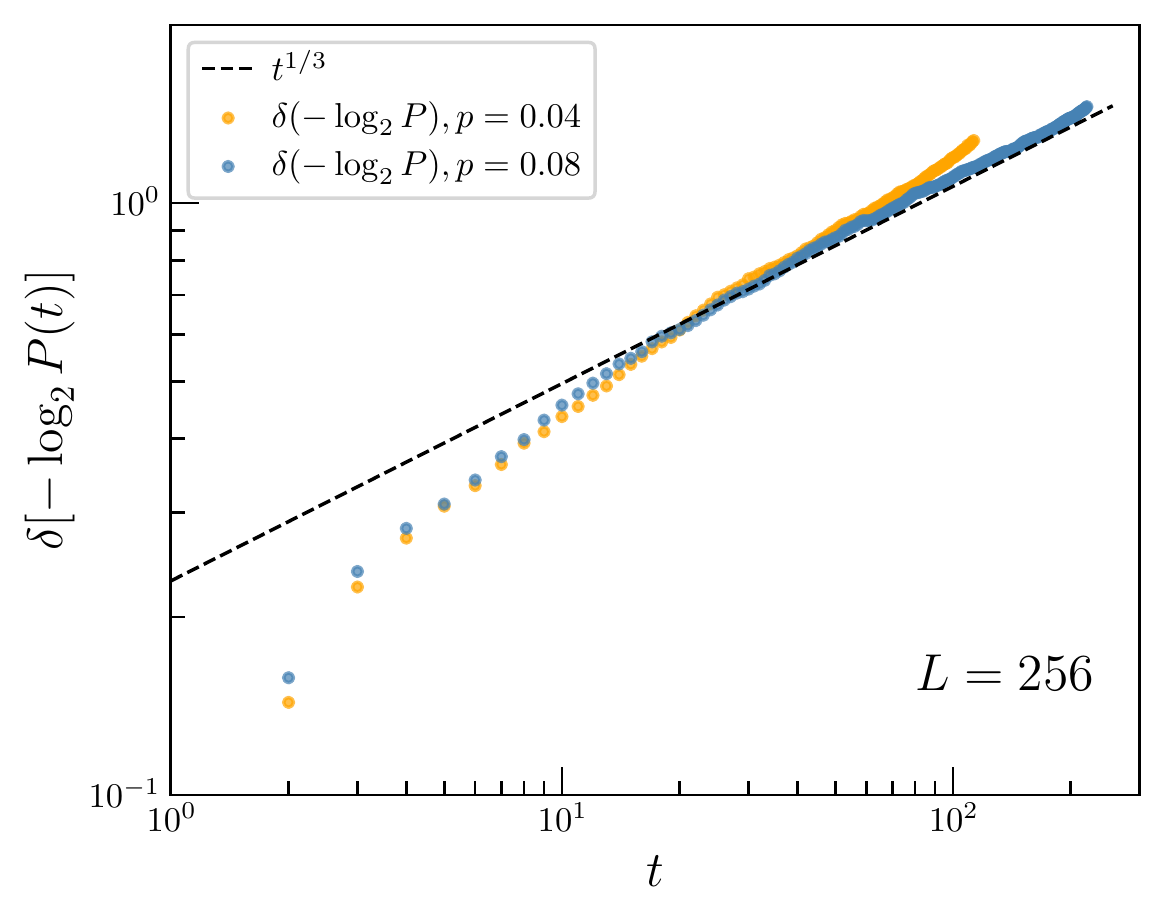}
    \label{fig:2ps_t}}
  \subfigure[]{
    \includegraphics[width=0.4\textwidth]{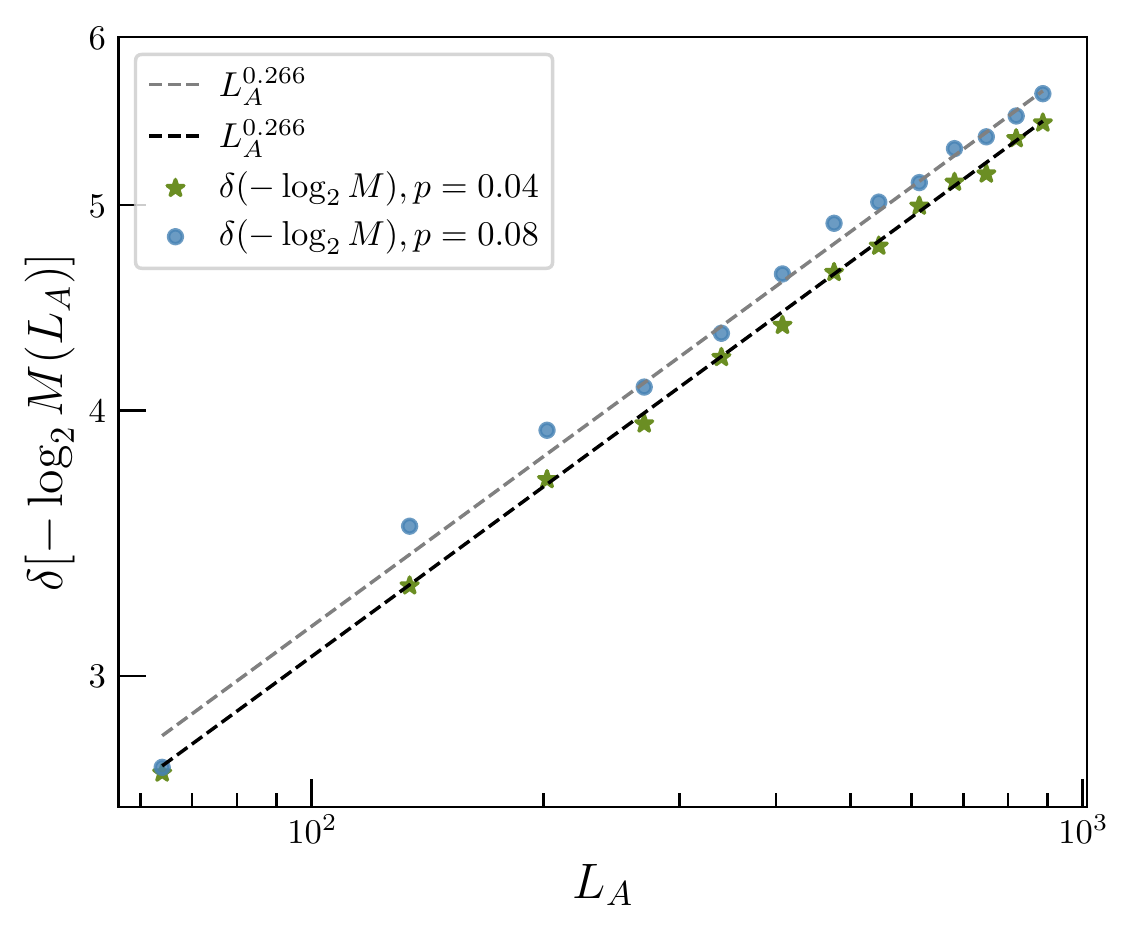}
    \label{fig:2ps_LA}}
  \caption{(a) The early time $\delta(-\log_2{P})$ vs $t$ plotted on a log-log scale. The numerical data are calculated from the two-species particle model with the sampling method and with the system size $L=256$. We find that $\delta[-\log_2{P(t)}] \propto t^{0.387}$ at $p=0.04$ and $t^{0.34}$ at $p=0.08$. (b) The steady state $\delta(-\log_2{M})$ vs $L_A$ plotted on a log-log scale, the ratio $L_A/L$ is fixed to be $1/2$. $M$ is one of the terms that contribute to $P$ which can be computed using the basis-decomposing method. }
\end{figure}

In the two-species particle model, it is unclear if there exists an efficient algorithm to evaluate $P(t)$. The existence of two moving fronts makes the problem difficult to solve. Nevertheless, we can still simulate the early time dynamics using Monte Carlo sampling method\cite{Iaconis_2021pmf}. More specifically, we prepare a large sample of randomly generated particle configurations and estimate $P(t)$ by computing the fraction of configurations in which $X$ and $Y$ never meet up to time $t$. This method works well for subsystem with entanglement entropy smaller than $\sim 20$. Around $p_c$, the entanglement entropy is small and this sampling method has been successfully used to identify $p_c$ and compute the critical exponents precisely over a few hundred qubits \cite{PhysRevB.105.064306}. However, deep in the volume-law phase, evaluating the sample fluctuation of $-\log_2{P(t)}$ is difficult since $P(t)$ soon becomes exponentially small and avoiding the contact between the two species turns into a rare event. Even though we prepare $O(10^7)$ number of particle configurations for each sample, the numerical $\overline{P(t)}$ decays to zero in a short time. Similarly, it is even more unrealistic for us to evaluate $\delta [-\log_2{P(L_A)}]$ in the steady state.

We apply the sampling method to the volume-law phase during the early time and as shown in Fig. \ref{fig:2ps_t}, the standard deviation of the entropy $\delta[-\log_2{P(t)}] \propto t^{0.34}$ at $p=0.08$, which is consistent with the KPZ fluctuation, and $t^{0.387}$ at $p=0.04$, which is already slightly off from $\beta=\frac{1}{3}$. For the steady state, we analyze the physics of $P$ below and take some approximation to estimate the fluctuation of $-\log_2 P$.

In the context of particle dynamics, the entanglement entropy saturates when all the particle configurations which contribute to $P(t)$ have at most one species left. The steady state $P$ is therefore composed of three parts,
\begin{equation}
    P=\frac{N_X}{2^L}+\frac{N_Y}{2^L}-\frac{N_{XY}}{2^L}=P_X+P_Y-P_{XY},
    \label{eq:ss}
\end{equation}
where $P_X$ ($P_Y$) denotes the fraction in which $X$ ($Y$) particles annihilate first under the dynamics before they could encounter the other species, $P_{XY}$ denotes the fraction in which both species extinguish  at the same time before they meet. In the volume-law phase, $P_X\propto \exp(-L_A)$, $P_Y\propto \exp(-L_{B})$ and $P_{XY}\propto \exp(-L)$. In the thermodynamic limit, the last term can be ignored and the first two terms compete as we tune $L_A$. When $L_A<L_{B}$, $P_X$ dominates and we have $P\approx P_X$. In contrast, when $L_A>L_{B}$, we have $P\approx P_Y$. This leads to 
\begin{equation}\label{eq:EE_ss}
  S_A^{(2)}\approx
  \begin{cases}
    -\log_2 P_X, & L_A< L/2 \\
    -\log_2 P_Y, & L_A >L/2.
  \end{cases}
\end{equation}

Computing $P_X$ is still not an easy task. Instead we consider a subset of $P_X$ that can be simulated efficiently using the basis-decomposing method in Sec.~\ref{sec: 1ps}. We define the binary basis $H_X^0$ ($H_Y^0$) which span the $X$ ($Y$) particle configurations in the absence of $Y$ ($X$) particles. Both $H_X^0$ and $H_Y^0$ evolve under the same dynamics. At time $t$, we consider the configurations in which the $X$ particles never encounter $Y$ particles in any of the basis states of $H_Y^0(t)$ and denote this fraction as $M(t)$. In other words, $M(t)$ is equivalent to $K(t)$ in the single-species particle model, except that now the boundary determined by the leftmost $Y$ particle in $H_Y^0(t)$ is spreading to the left. Therefore,
\begin{equation}
    -\log_2{M(t)} =-\log_2{\frac{2^{|H_X(t)|}\times 2^{L_{B}}}{2^L}}=L_A-|H_X(t)|.
\end{equation}
where $H_X(t)$ is the basis of $X$ particle configurations which never meet the leftmost $Y$ particle in $H^0_Y(t)$.

\begin{table}
\begin{tabular}{|c|c|c|c|c|}
    \hline
    \multicolumn{2}{|c|}{ } & $p=0$ & $p=0.04$ & $p=0.08$ \\
    \hline
    \multirow{2}{*}{$\delta S_A$} & $\beta_1$ & N/A & 0.322 & 0.31 \\
    & $\beta_2$ & 0.307 & 0.307 & 0.266 \\
    \hline
    \multirow{2}{*}{$\delta(-\log_2{K})$} & $\beta_1$ & N/A & 0.245 & 0.245 \\
    & $\beta_2$ & 0.304 & 0.294 & 0.26 \\
    \hline
    $\delta(-\log_2{P})$ & $\beta_2$ & N/A & 0.34 & 0.387 \\
    \hline
    $\delta(-\log_2{M})$ & $\beta_1$ & N/A & 0.266 & 0.266 \\
    \hline
\end{tabular}
\caption{The comparison of the exponents of the fluctuation $\delta S_A$ of Clifford QA entanglement entropy, $\delta(-\log_2{K})$ of the single-species particle model, the early time $\delta[-\log_2{P(t)}]$ of the two-species particle model using the sampling method, the steady-state $\delta[-\log_2{M(L_A)}]$ where $M$ is a term that contributes to $P$. The measurement rate or the particle annihilation rate is taken to be $p=0, 0.04$ and $0.08$. In the table, $\beta_1$ refers to the exponent of $L_A^{\beta_1}$ and $\beta_2$ refers to the exponent of $t^{\beta_2}$.}
\label{table:exponents}
\end{table}

In the steady state, $H_X$ is the set of basis in which the $X$ particles have already vanished before encountering any $Y$ particles in $H_Y^0(t)$. As shown in Fig. \ref{fig:2ps_LA}, there exists sample fluctuation in $-\log_2{M}$ in the volume-law phase. In particular, we find that  $\delta[-\log_2{M}]\propto L_A^{0.266}$ for both $p=0.04$ and $p=0.08$. This exponent is smaller than the one computed in the Clifford QA circuit. The exponents for different models at various $p<p_c$ are listed in Table. \ref{table:exponents} and we find that some of them are smaller than $1/3$.


Currently, it is unclear if this is a finite size effect, or if the fluctuations of these quantities in the one/two-species particle models belong to other universality classes. The main obstacle of this issue is the lack of rigorous analytical results. However, we want to mention that there are some known results about KPZ fluctuations in the particle dynamics. Under the hybrid QA circuit, each particle configuration experiences the same circuit dynamics, therefore the end points of the two species $x$ and $y$ can be treated as particles performing biased random walks in a fixed time-dependent random environment. Mathematically, the dynamics of the endpoint is known as random walk in random environment (RWRE), in which the logarithm of the transition probability is proven to exhibit KPZ fluctuations in some limit \cite{Corwin_2017wa,Barraquand_2017vr,PhysRevE.96.010102}. Indeed, this quantity is similar to the second R\'enyi entropy and the detailed discussion about this connection can be found in Appendix~\ref{Appendix:RWRE}.

\section{Purification Process and quantum error correction}
\subsection{Purification process and hybrid QA QECC}\label{sec: QA_puri}

 \begin{figure}[htp!]
  \centering
  \subfigure[]{
    \includegraphics[width=0.2\textwidth]{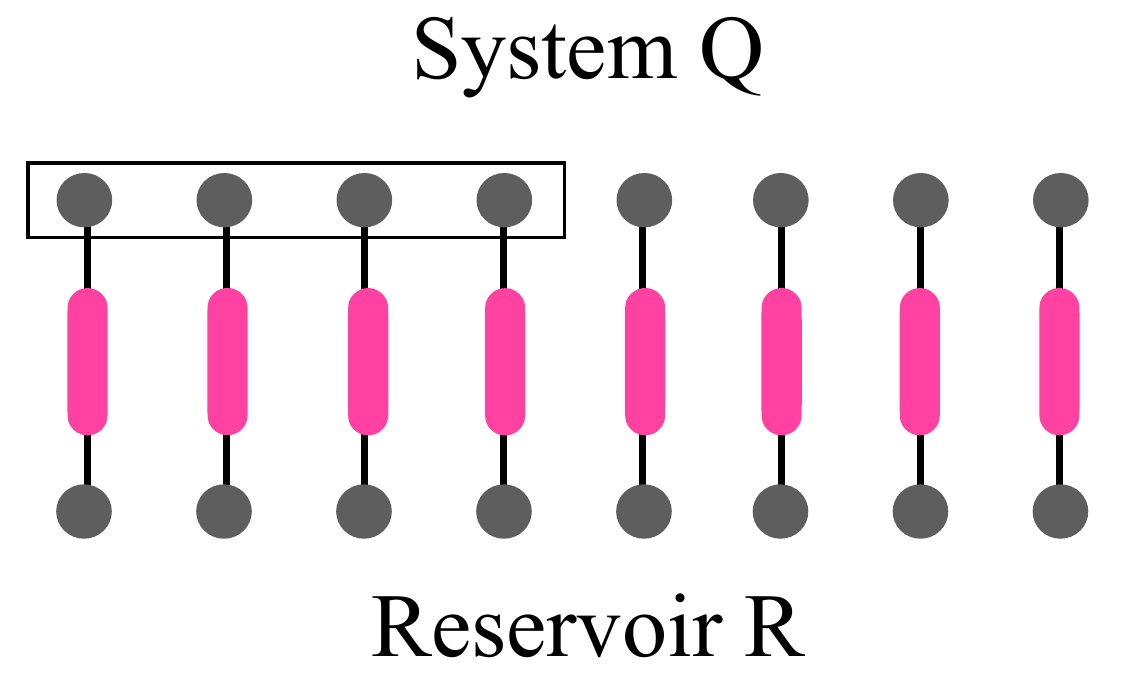}
    \label{fig:Phase_gate}
  }
  \subfigure[]{
    \includegraphics[width=.3\textwidth]{EE_gates.pdf}
    \label{fig:Puri_gates}
  }
  \subfigure[]{
        \includegraphics[width=.3\textwidth]{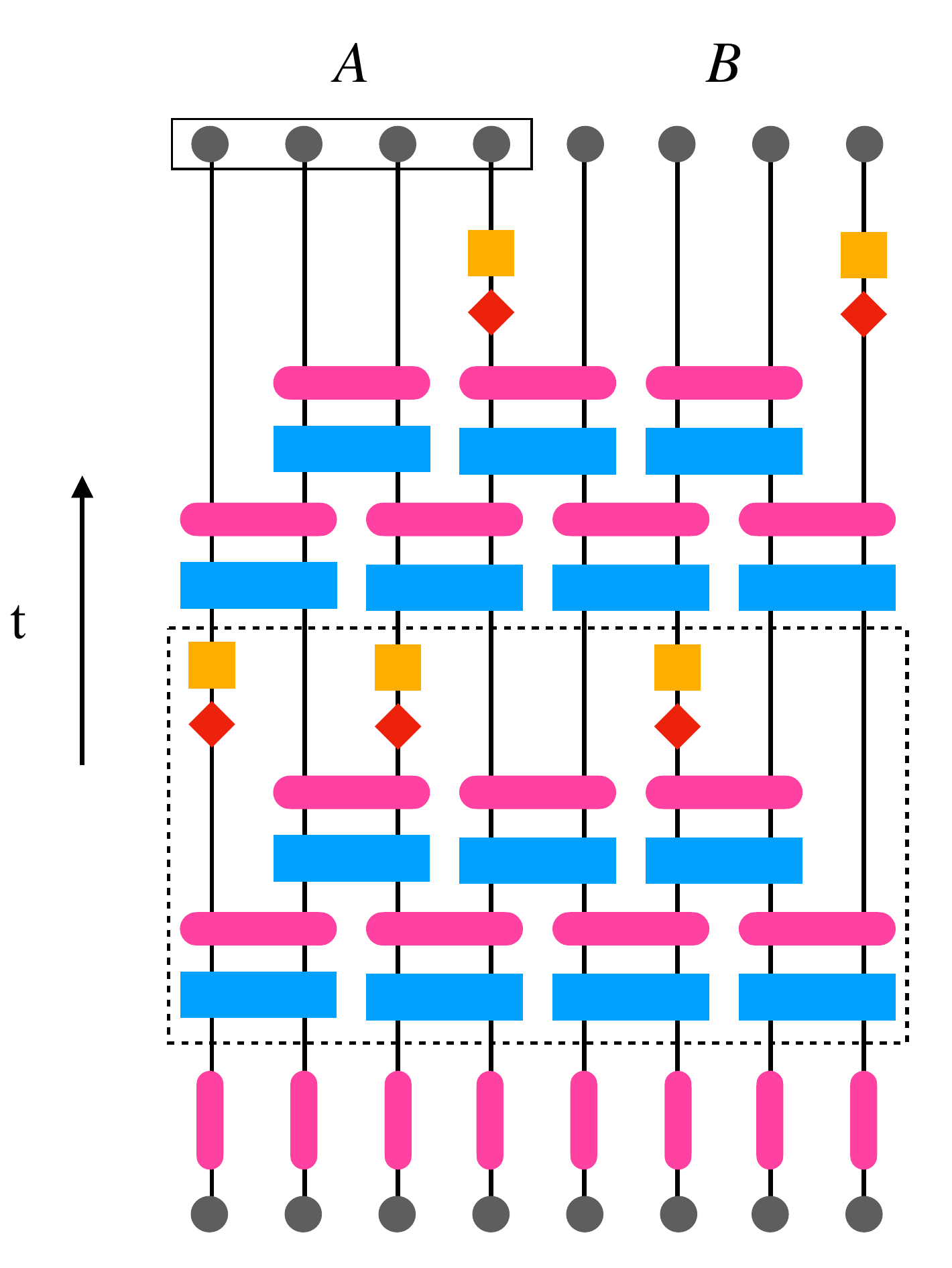}
    \label{fig:QA_puri}
  }

  \caption{(a) We use CZ gates to generate entanglement between system $Q$ and environment $R$. (b) The symbols of the CZ gate, CNOT gate, the single-qubit Z measurement gate and Hadamard gate. (c) The arrangement of gates for the purification process of the hybrid QA circuit. Except the initial setup in (a), the hybrid circuit is applied solely in system $Q$.}
  \label{fig:purification}
  \end{figure}
  
An alternative approach to understand the measurement-induced entanglement phase transition is through purification dynamics\cite{PhysRevX.10.041020}. The basic idea is to prepare a system $Q$ with an extensive entropy and evolve it under the hybrid quantum dynamics. Although the system will eventually be purified, in the weakly monitored volume-law phase with $0<p<p_c$, the purification time is exponentially long in system size $L$. On the other hand, when $p>p_c$, the entropy decays exponentially in time with a constant rate. 

The existence of long purification time in the regime $0<p<p_c$ suggests that the hybrid quantum circuits can dynamically generate a robust quantum error correcting code (QECC) at polynomial time\cite{PhysRevX.10.041020}. The QECC can be compactly denoted as $[L,k,d]$. Here $L$ is the number of physical qubits in $Q$ and $k$ characterizes the amount of logical information encoded in the code space and is quantified by the entropy of $\rho_Q$. The third index $d$ is the code distance,  which is defined as the minimum weight of all uncorrectable errors. Here the errors can be interpreted as measurements which can potentially reduce the entropy when applied on $\rho_Q$. If an error changes the entropy of $Q$, then it is an uncorrectable error since it damages the encoded quantum information and hence can not be recovered. Due to the locality of the circuit model, a better measure is the contiguous code distance $d_{\text{cont}}$, which is the minimal length of a contiguous section of qubits that supports an uncorrectable error \cite{PhysRevX.10.041020}. In the volume-law phase, $d_{\text{cont}}$ diverges in the thermodynamic limit. The quantum information is stored non-locally under the unitary evolution and thus is protected against any local measurements. On the other hand, for $p>p_c$, the unitary dynamics fails to protect the encoded information under frequent measurements. Previous works have quantitatively analyzed the statistical property of QECC in hybrid random Clifford circuits \cite{PhysRevB.103.104306, li2021entanglement}. In this section, we will study the purification dynamics of the hybrid QA circuit and investigate the QECC in terms of the classical particle model.

\begin{figure}[tp!]
  \centering
  \subfigure[]{
    \includegraphics[width=0.4\textwidth]{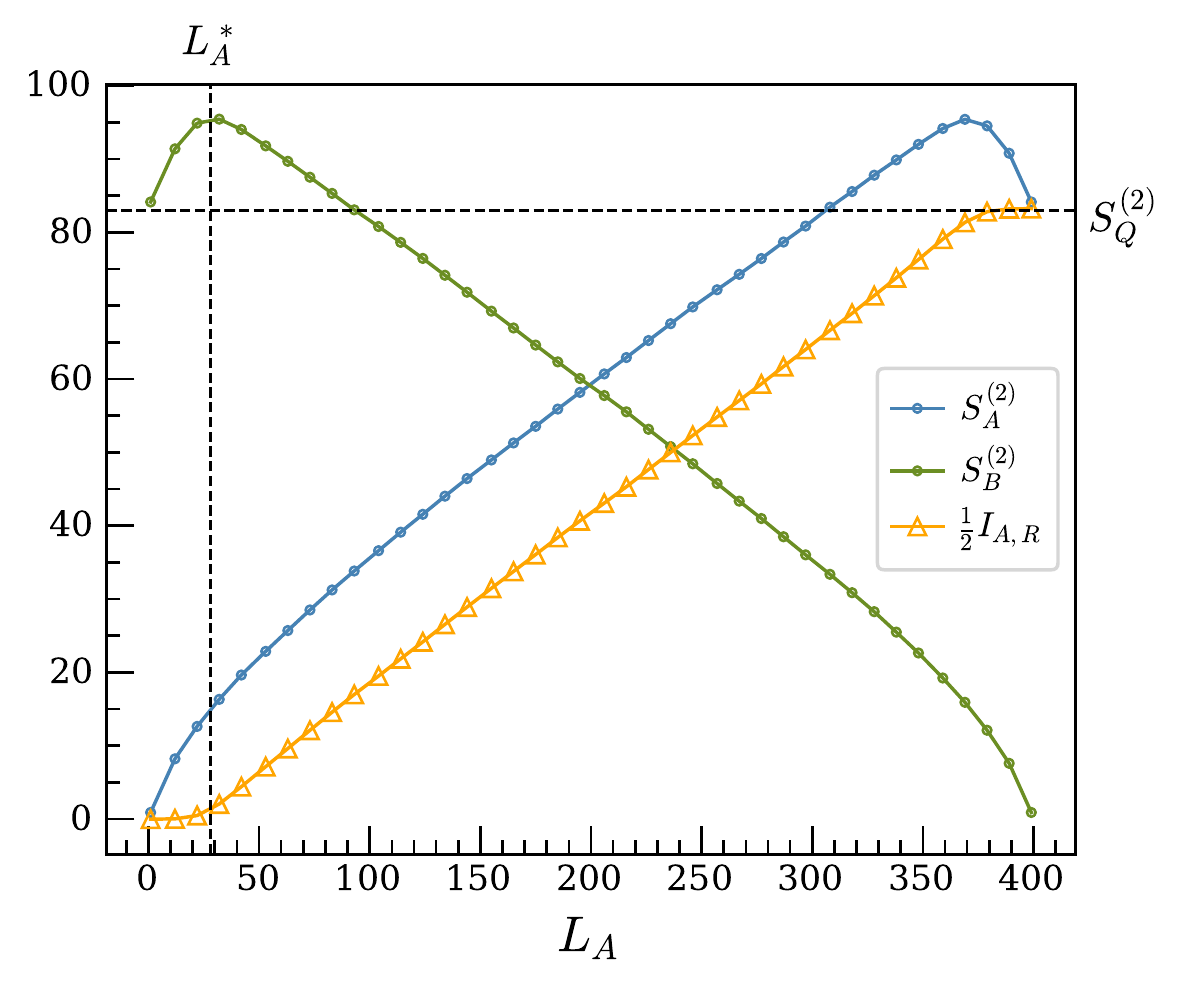}
    \label{fig:puri_I_AR}
  }
  \subfigure[]{
    \includegraphics[width=0.4\textwidth]{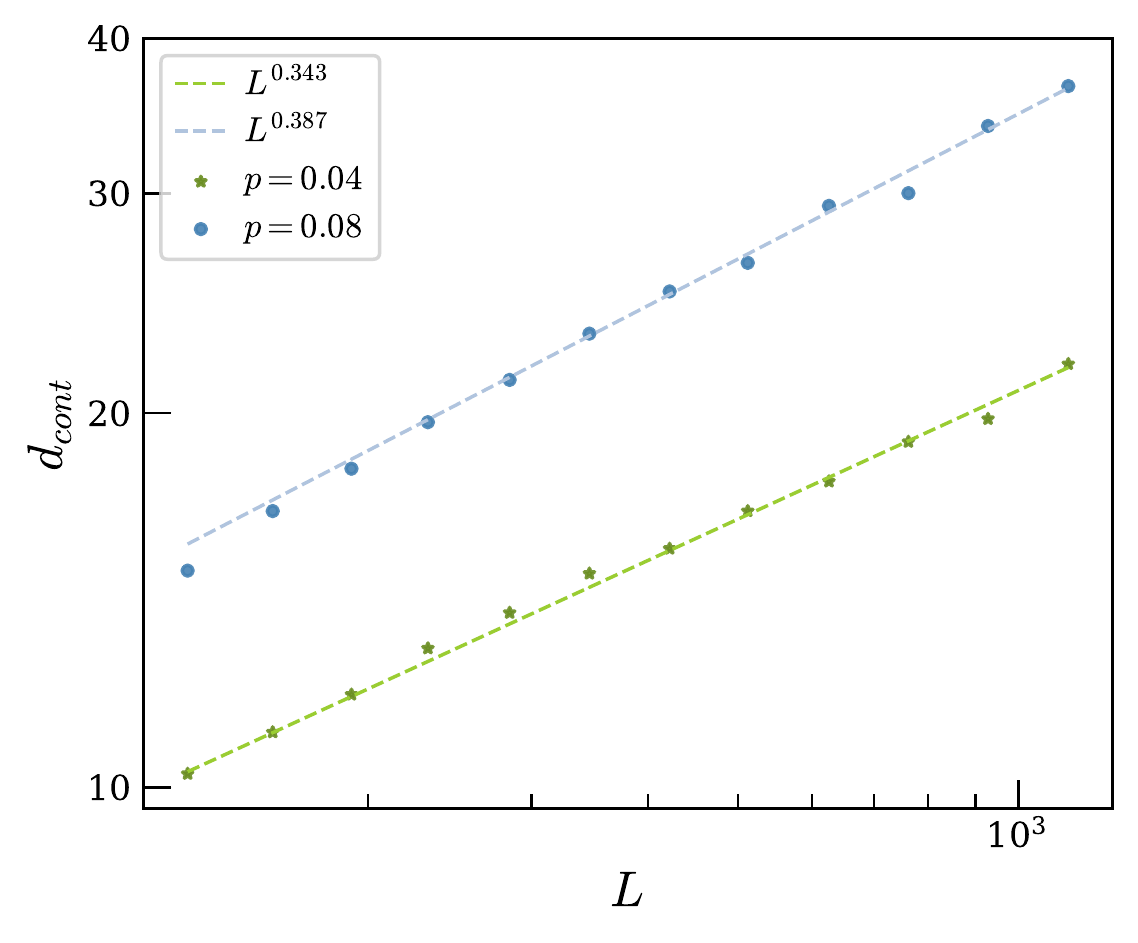}
    \label{fig:puri_d}
  }

  \caption{(a) The entanglement entropies $S^{(2)}_A$ and $S^{(2)}_{B}$ and half of the mutual information $\frac{1}{2}I_{A,R}$ vs $L_A$ computed from the Clifford QA circuit with system size $L=400$ at $T=2L$, and $p=0.08$. (b) The contiguous code distance $d_{\text{cont}}$ for different system size $L$ at $p=0.08$ and $p=0.04$ at $T=3L$ plotted on a log-log scale. Here we take $d_{\text{cont}}$ to be the maximum length $L_A^*$ such that $\langle I_{A,R} \rangle \leq 1$. All of the data are computed with PBC.}
  \label{fig:code_dis}
\end{figure}

Initially, we prepare a product state of $2L$ qubits polarized in $+x$ direction which can be evenly divided into system $Q$ and environment $R$. Then, we create $L$ EPR pairs between them by applying CZ gates as shown in Fig. \ref{fig:Phase_gate}. Thus, the system $Q$ becomes maximally entangled with environment $R$, i.e., $S_Q^{(2)}=L$. To investigate the purification dynamics, a hybrid QA circuit is solely applied on system $Q$. Numerically, we consider the model described in Fig. \ref{fig:QA_puri}, identical to the hybrid QA Clifford circuit in the entanglement dynamics in the last section. The composite measurements disentangle the qubits in $Q$ from the environment $R$. Meanwhile, the unitary evolution scramble the quantum information within system $Q$, increasing the entanglement between any subsystem $A$ in $Q$ and its complement $B:= Q\setminus A$, but not affecting $S_Q^{(2)}$. There is a phase transition in the purification time of $S_Q^{(2)}$ at $p=p_c\approx 0.138$ \cite{Iaconis_2021pmf}, consistent with the entanglement dynamics studied in the last section.

As we discussed before, an important measure of the error-correcting ability of our QECC is the contiguous code distance $d_{\text{cont}}$, which is the minimal length of a contiguous region supporting an uncorrectable error. It is thus natural to mark the existence of such errors supported on a contiguous subregion $A$ using the mutual information between $A$ and the environment $R$ \cite{Hayden_2008,PhysRevB.103.104306,Brown_2015},
\begin{equation}\label{eq:MI_AR}
  \begin{aligned}
    I_{A,R}&=S^{(2)}_A+S^{(2)}_R-S^{(2)}_{A,R}
    \\
    &=S^{(2)}_A+S^{(2)}_Q-S^{(2)}_{B}.
  \end{aligned}
\end{equation}
When $I_{A,R}=0$, $A$ and $R$ are completely decoupled, we cannot acquire any information encoded in the state by observing any qubits within the subregion $A$. In other words, any  measurements acting within $A$ are correctable errors since they will not affect $S^{(2)}_Q$. Therefore, $d_{\text{cont}}$ is the maximum length $L_A^*$ such that $I_{A,R}=0$ for $L_A<L_A^*$ \footnote{In the stabilizer QECC, $d_{\text{cont}}$ can also be viewed as the minimal length of nontrivial logical operators acting within the code space. This is equivalent to the $d_{\text{cont}}$ defined through the criterion $I_{A,R}=0$.}.

We simulate the Clifford QA circuit to find the entanglement entropies $S^{(2)}_A$ and $S^{(2)}_{B}$ and the mutual information $I_{A,R}$ over various subsystem sizes $L_A$. The numerical results are given in Fig. \ref{fig:code_dis}. We take the code distance to be the maximum length $L_A^*$ such that $\langle I_{A,R}\rangle \leq\epsilon$ for $L_A<L_A^*$. In the numerical simulation of the finite system size, we set $\epsilon=1$. Remarkably, we find that $S^{(2)}_A$ starts to decrease at $L_A=L-L_A^*$ until it reaches $S^{(2)}_Q$ at $L_A=L$. This non-monotonic behavior coincides with that in the previous study of the hybrid Clifford circuits \cite{PhysRevB.103.104306} and is crucial in understanding the code distance. We will modify the two-species particle model in the following section so as to give an interpretation for $S^{(2)}_A$. As shown in Fig. \ref{fig:puri_d}, $d_{\text{cont}}$ has a sublinear power law scaling with $L$. Numerically, it scales as $L^{0.343}$ for $p=0.04$ and $L^{0.387}$ for $p=0.08$, and its value increases as the measurement rate increases.

\subsection{QECC in classical particle language}

To understand the dynamically generated QECC from the perspective of classical particle dynamics, we need to compute the mutual information defined in Eq.~\eqref{eq:MI_AR} in terms of the two-species particle model. An important task is to understand the entanglement entropy of a subsystem $A$ in the presence of environment $R$. For the bit-string dynamics in the purification process, the hybrid QA circuit is applied only on system $Q$ of the bit-strings in a time-reversed order, generating the relative phase $\Theta_r$, followed by the CZ gates acting on both the system $Q$ and environment $R$, generating another relative phase $\Delta_r$. Therefore, only the configurations satisfying $\Theta_r=0$ and $\Delta_r=0$ contribute to the purity. As shown in Appendix \ref{Appendix:2ps_puri}, in the particle picture, this corresponds to the configurations in which all of the $X$ particles have vanished before they can encounter any $Y$ particles at time $t$. These configurations are a subset of $N(t)$ defined in Eq.\eqref{eq:dyn} in the entanglement dynamics. Let the number of these configurations be $N_1(t)$, the entanglement entropy of $A$ is then
\begin{equation}\label{eq:puri_bit_EE}
  S_A^{(2)}(t)=-\log_2\frac{N_1(t)}{2^L}\equiv -\log_2{P_1(t)}.
\end{equation}
Specifically, when $A=Q$, there are only one type of particles, we only need to count the configurations whose particles extinguish at time $t$. Letting the number of such configurations be denoted $N_Q$, we have
\begin{equation}
    S_Q^{(2)}(t)=-\log_2\frac{N_Q(t)}{2^L}\equiv-\log_2 P_Q(t).
\end{equation}
Initially, $P_Q(t=0)=1/2^L$ and $Q$ is maximally entangled with $R$. Under the hybrid QA circuit, more and more configurations become empty and $S_Q$ decreases monotonically with time. The time scale for which the particles of all the configurations vanish depends on $p$ and is consistent with that of the purification transition.

We are interested in the QECC generated at polynomial time $t=\lambda L$ with $\lambda\gg 1$. At this time, $X$ or $Y$ particles have already spread over the entire system and therefore the configurations that contribute to $P_1(t)$ can have at most one type of particle. Similar to the steady state $P$ of the entanglement dynamics derived in Eq.\eqref{eq:ss}, $P_1(t)$ can be expressed as 
\begin{equation}
    P_1(t)=\frac{\widetilde{N}_X(t)}{2^L}+\frac{\widetilde{N}_Y(t)}{2^L}
    \equiv P_X+\widetilde{P}_Y(t),
\end{equation}
where $\widetilde{P}_Y(t)$ is a subset of $P_Y$, which further requires that $X$ particles vanish at time $t$. There is also a small contribution from $P_{XY}$ which we ignore here.

\begin{figure}[tp!]
  \centering
  \subfigure[]{
  \includegraphics[width=0.4\textwidth]{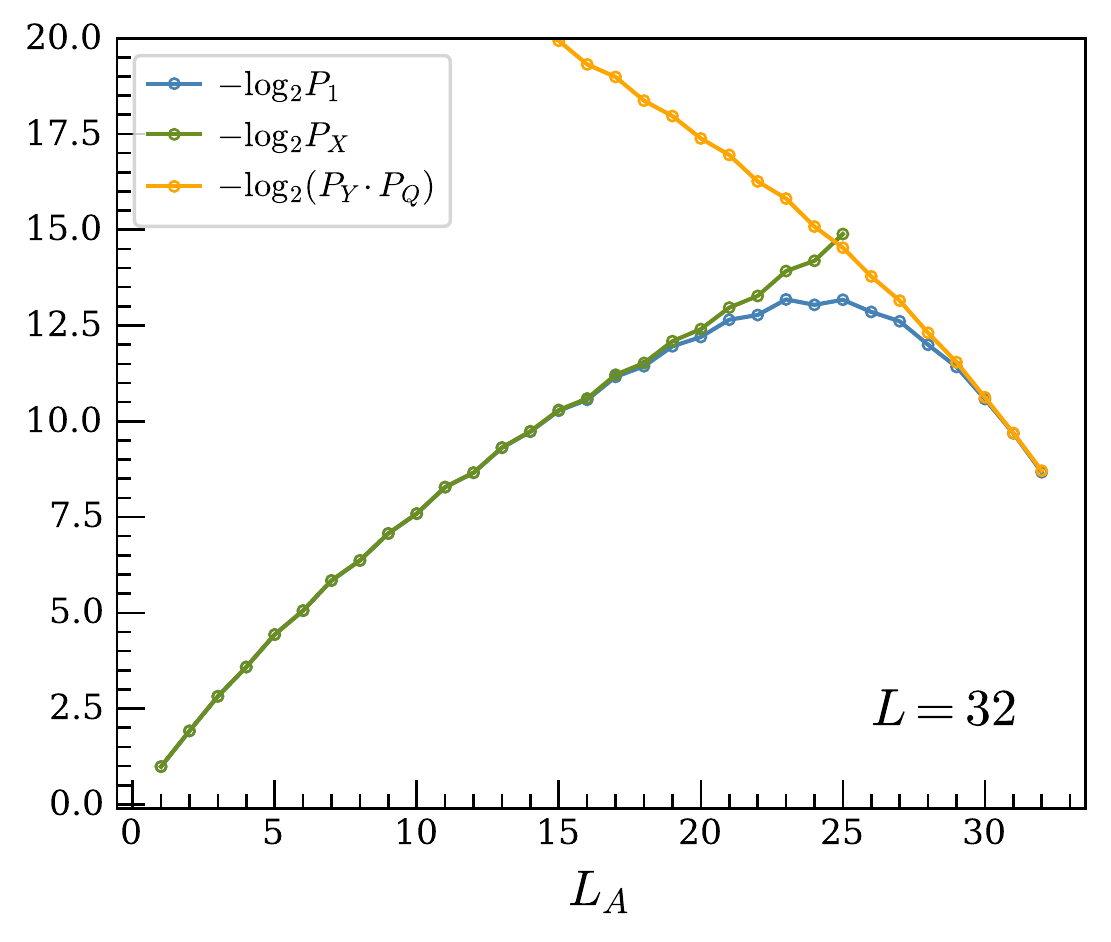}
  \label{fig:puri_bit_S}}

  \subfigure[]{
  \includegraphics[width=0.4\textwidth]{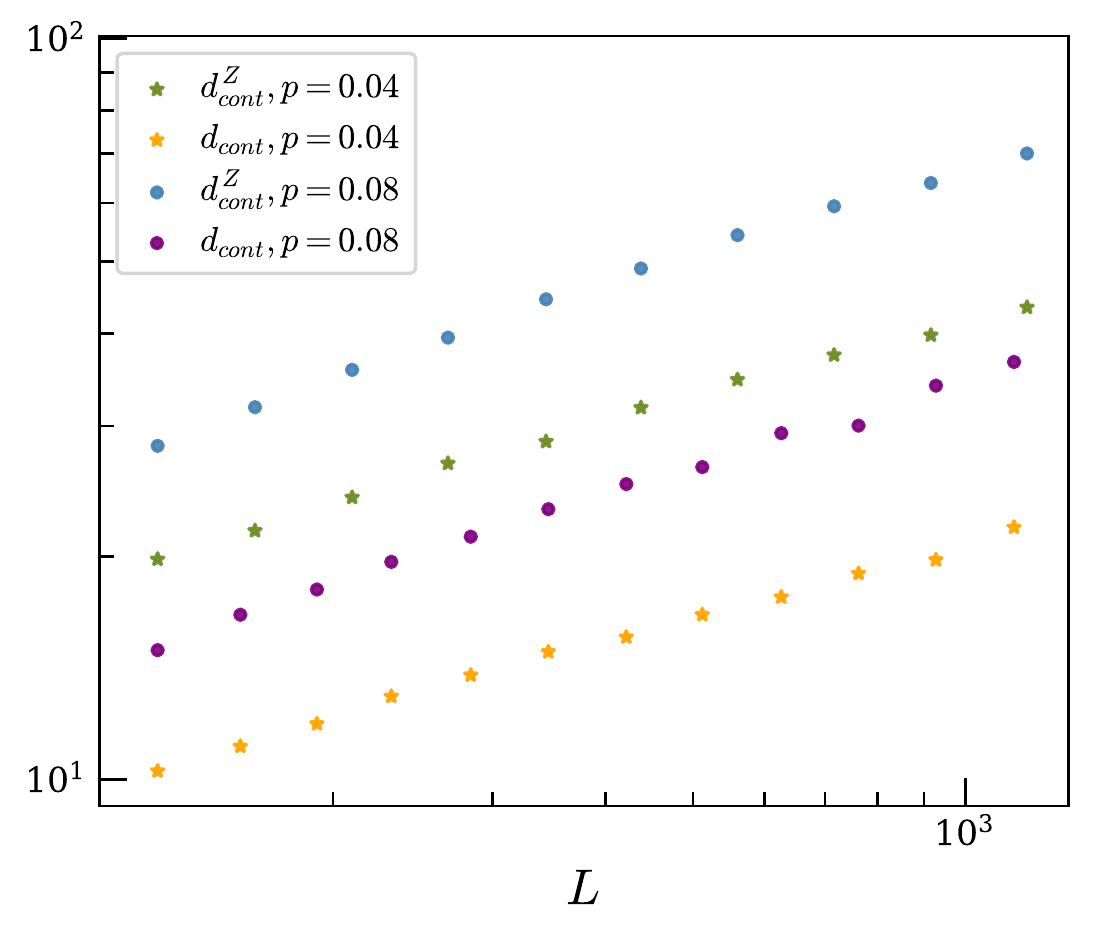}
  \label{fig:particle_Z_d}}

  \caption{(a) The entanglement entropy $S_A^{(2)}\approx -\log_2{P_1}$ vs $L_A$ computed from the two-species particle model, in comparison with the two approximate values $-\log_2{P_X}$ and $-\log_2{(P_Y\cdot P_Q)}$ for $L_A< L^c$ and $L_A> L^c$. The numerical data are calculated by the sampling method over a system of size $L=32$, at $T=3L$, $p=0.08$ and under PBC. (b) The $Z$-error contiguous code distance $d^Z_{\text{cont}}$ in comparison with $d_{\text{cont}}$ for different system sizes at $p=0.04$ and $p=0.08$ at $T=3L$ plotted on a log-log scale. We find that $d^Z_{\text{cont}}$ scales as $L^{0.327}$ when $p=0.04$ and $L^{0.366}$ when $p=0.08$.}
\end{figure}

When $L_A<L_B$, since $P_X>P_Y>\widetilde{P}_Y(t)$, $P_X$ dominates and we have $P_1(t)\approx P_X$. Therefore, $S_A^{(2)}(t)$ is the same as the steady state $S_A^{(2)}$ in the entanglement dynamics. The regime $L_A>L_B$ is different from that of the steady state in the entanglement dynamics. Since $\widetilde{P}_Y(t)$ is a small fraction of $P_Y$, when $L_A$ is slightly larger than $L_B$, $P_X> \widetilde{P}_Y(t)$ and we still have $S_A^{(2)}\approx -\log_2 P_X$. We define $L^c$ to be the threshold of the subsystem size $L_A$ where $P_X=\widetilde{P}_Y(t)$. When $L_A>L^c$, $\widetilde{P}_Y(t)$ dominates and we have $S_A^{(2)}(t)\approx -\log_2 \widetilde{P}_Y(t)$. For $\widetilde{P}_Y(t)$, it can be understood as follows,
\begin{align}
    \widetilde{P}_{Y}(t)=\frac{\widetilde{N}_{Y}(t)}{2^{L}}=\frac{N_{Y}}{2^L}\frac{\widetilde{N}_{Y}(t)}{N_{Y}}\equiv P_{Y}P_2(t),
\end{align}
where $P_2(t)\equiv \widetilde{N}_Y(t)/N_Y$. Since the $X$ particles of configurations in $N_Y$ have already spread over the entire system when all of the $Y$ particles extinguish, $P_2(t)$ actually counts the fraction of configurations which have no particles at time $t$. Directly evaluating $P_2(t)$ is difficult. However, due to the scrambling property of the unitaries, it is reasonable to assume that $P_2(t)\approx P_Q(t)$, 
\begin{align}
    \widetilde{P}_Y(t)\approx P_Y P_Q(t).
\end{align}
Summarizing, we have
\begin{equation}\label{eq:puri_S_A}
  S_A^{(2)}(t)\approx
  \begin{cases}
    -\log_2 P_X, & L_A< L^c \\
    -\log_2 P_{Y}-\log_2 P_Q(t), & L_A>L^c.
  \end{cases}
\end{equation}
We numerically verify the above approximation in Fig. \ref{fig:puri_bit_S}. Due to the difficulty for simulating highly-entangled state, we are only able to simulate the two-species particle model on a system with size $L=32$ at $T=3L$ in the volume-law phase. We find that there indeed exists a non-monotonic decay area for $S_A^{(2)}=-\log_2{P_1}$ when $L_A$ passes the threshold $L^c$, and that Eq.\eqref{eq:puri_S_A} holds within a small difference $\epsilon=O(1)$. The non-monotonicity comes from the competition of the two terms $P_X$ and $\widetilde{P}_Y$. As subsystem $A$ enlarges, $P_Y$ increases and $P_Q$ stays the same. As a result, when $L_A>L^c$ and $\widetilde{P}_Y$ dominates, $S_A^{(2)}$ starts to decline as $L_A$ continues to increase. The location of the peak $L^c$ depends on time and can eventually shift to $L/2$ when the system is completely purified.

Based on the above analysis of $S_A^{(2)}(t)$, we are now ready to understand the QECC in terms of particle dynamics. In the regime with $L_A\in[0,L-L^c)$ and hence $L_B\in(L^c,L]$, 
the mutual information becomes
\begin{equation}
    \begin{aligned}
        I_{A,R}&=S_A^{(2)}+S_Q^{(2)}-S_B^{(2)} \\
        &\approx -\log_2 P_X-\log_2 P_Q+\log_2 P_X+\log_2 P_Q \\
        &=0.
    \end{aligned}
\end{equation}
It vanishes because the two terms in $S_B^{(2)}$ completely cancel with $S_A^{(2)}$ and $S_Q^{(2)}$, similar to the decoupling domain wall picture discussed in Ref.~\onlinecite{PhysRevB.103.104306}. On the other hand, when $L^c>L_A>L-L^c$, it is easy to show that \begin{equation}
    I_{A,R}\approx -\log_2 P_X -\log_2 P_Q +\log_2 P_Y >0.
\end{equation}
We arrive at the conclusion that $I_{A,R}=0$ \emph{if and only if} $L_A<L-L^c$ and the contiguous code distance is $d_{\text{cont}}=L-L^c$. These results are consistent with the numerical results of hybrid Clifford QA circuit in Fig.~\ref{fig:puri_I_AR}.

The code distance specified by the mutual information works for \emph{all} kinds of errors. In the QA circuit, we could consider a special type of error which is the $Z$ error defined as the measurement operator $(1\pm O)/2$ where $O$ is a Pauli $Z$ string. Suppose at time $t$, a QECC is prepared through the QA purification dynamics and some $Z$ errors occur within a contiguous subsystem $A$, 
which could possibly reduce the entropy of $\rho_Q$. We define the $Z$-error contiguous code distance $d^Z_{\text{cont}}$ as the maximum length of subsystem $A$ such that $S_Q^{(2)}$ does not change. 

Since the particle dynamics is evolved in a time-reversed order, the $Z$ error acts as annihilation on all of the particle configurations at $t=0$. For subsystem $A$ of size smaller than $d^Z_{\text{cont}}$, $S_Q^{(2)}$ is invariant under any $Z$ error occurred within $A$, or in other words, with any initial particle distribution in $A$. Therefore, we start from an ensemble of particle configurations with \emph{empty} subsystem $A$, so that the information about $A$ is completely removed. The entanglement entropy of $Q$ after the $Z$ error becomes
\begin{equation}
  S_Q^{(2)}(t)= -\log_2{P_{B}(t)},
\end{equation} 
where $P_B(t)$ denotes that among all the configurations with only $Y$ particles located in $B$ initially, the fraction that becomes completely empty at time $t$. Consequently, $d^Z_{\text{cont}}$ is the maximum length of subsystem $A$ such that $-\log_2{P_{B}(t)}=-\log_2{P_{Q}(t)}$ for $L_A<d^Z_{\text{cont}}$.

Both $-\log_2{P_{B}(t)}$ and $-\log_2{P_{Q}(t)}$ can be efficiently calculated by evaluating the number of independent basis under the hybrid time evolution. We prepare two sets of binary basis, one is $H$ whose rows are the basis spanning all the particle configurations in system $Q$, the other one is $H'$ which is obtained by replacing a contiguous submatrix of size $L\times L_A$ from $H$ with $0$'s. Then, we evolve them under the same circuit dynamics. One can easily see that $-\log_2{P_{Q}(t)}=\text{rank}_2(H(t))$ and $-\log_2{P_{B}(t)}=\text{rank}_2(H'(t))$, which are the number of independent basis in $H(t)$ and $H'(t)$ respectively. The code distance $d^Z_{\text{cont}}(t)$ is therefore identified as the largest $L_A$ such that the rank of $H(t)$ and $H'(t)$ agree within $\epsilon=1$. As shown in Fig. \ref{fig:particle_Z_d}, although $d^Z_{\text{cont}}$ is much larger than $d_{\text{cont}}$, they have similar power-law scaling.

\begin{figure}[tp!]
  \centering
  \subfigure[]{
    \includegraphics[width=0.3\textwidth]{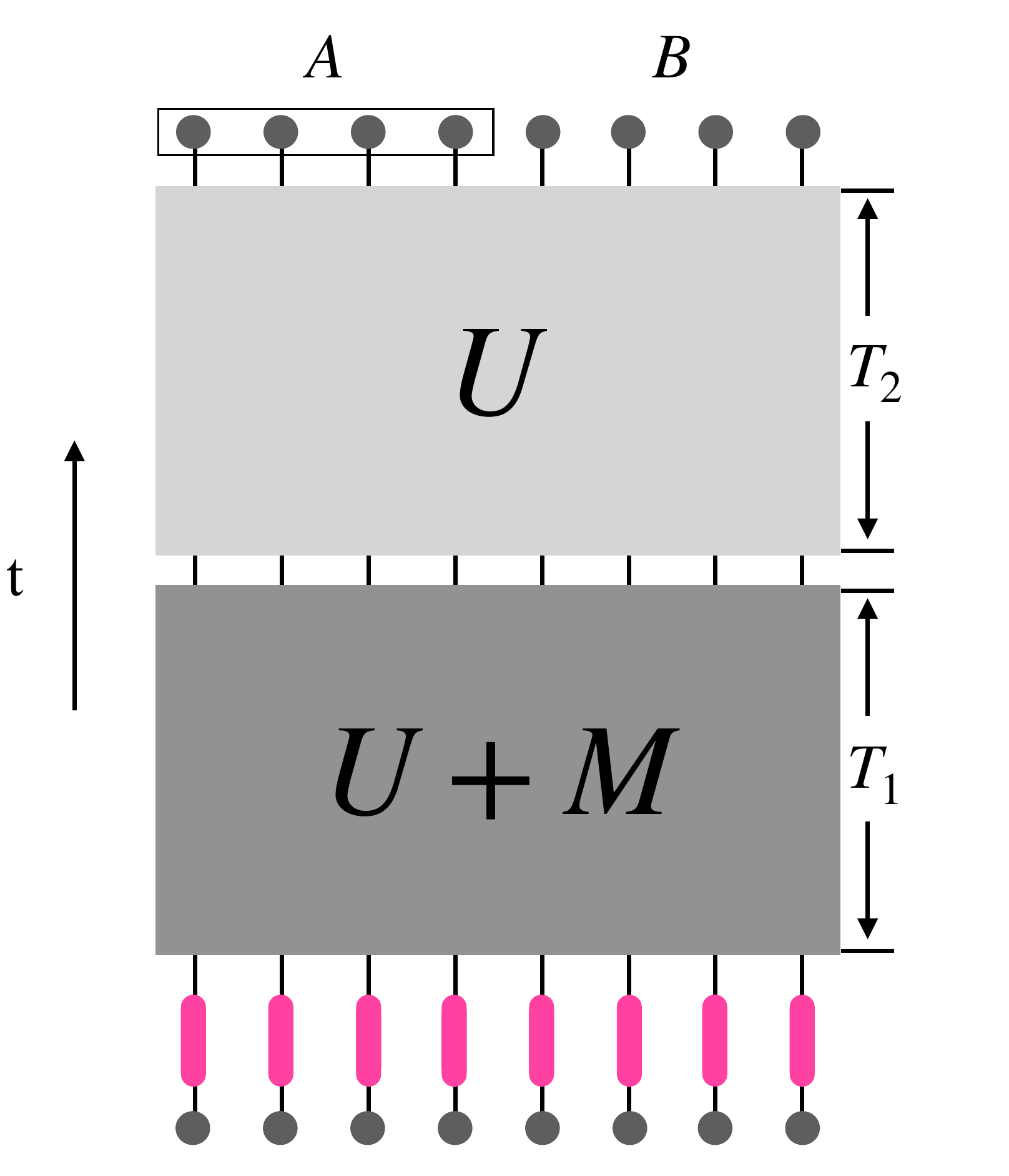}
    \label{fig:UP_U_circuit}
  }
  \subfigure[]{
    \includegraphics[width=0.45\textwidth]{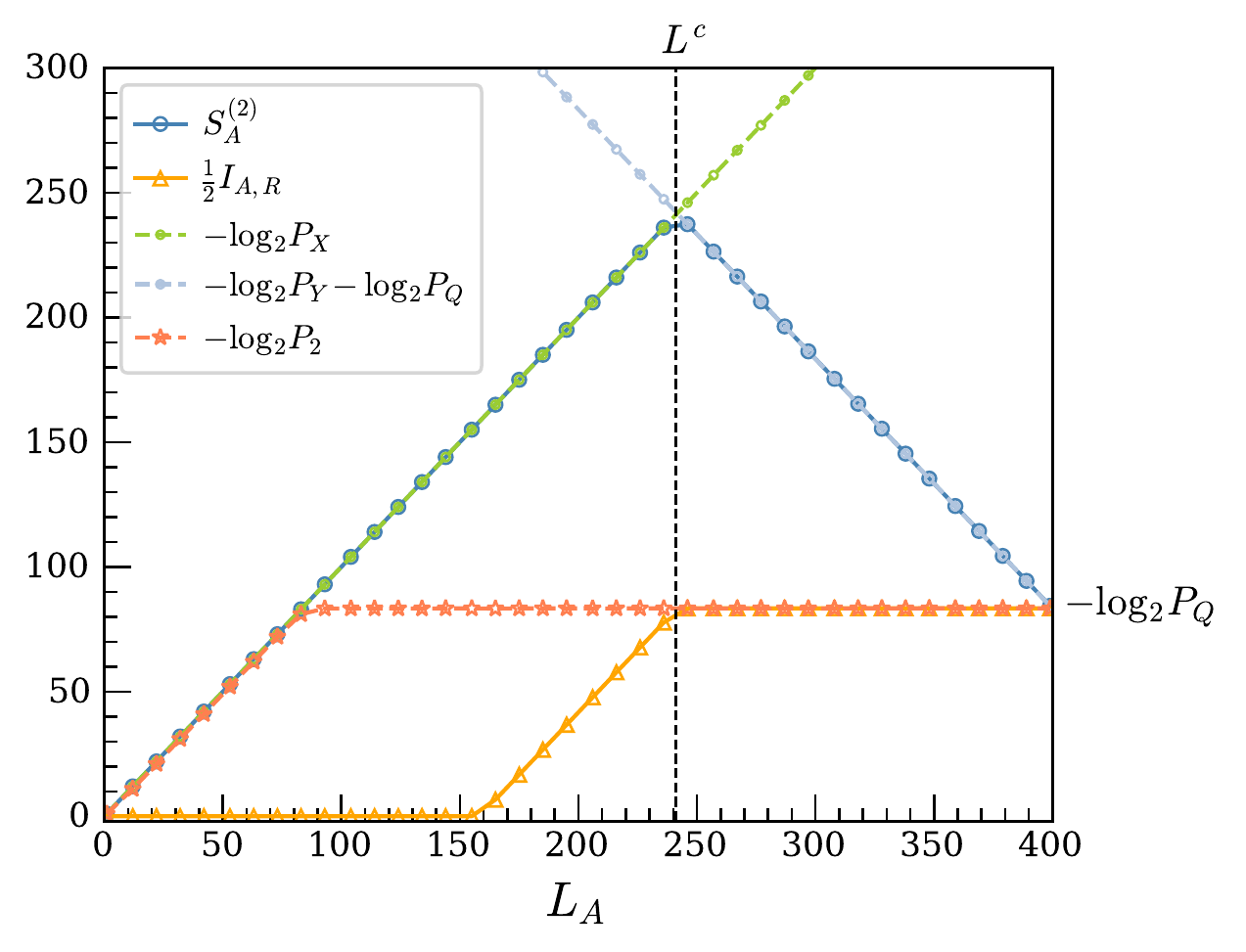}
    \label{fig:UP_U_comparison}
  }

  \caption{(a) The setup of the alternative hybrid QA circuit. ``U+M'' represents the original hybrid QA circuit composed of unitary gates and sporadic local composite measurements. ``U'' represents the circuit in the limit $p=0$ with only QA unitaries. (b) The entanglement entropy $S_A^{(2)}$ and half of the mutual information $\frac{1}{2}I_{A,R}$ vs $L_A$ computed from the Clifford model, in comparison with the two approximations $-\log_2 P_X$ and $-\log_2(P_Y\cdot P_Q)$. We also calculate $-\log_2 P_2$ and find that it grows linearly in $L_A$ and saturates to $-\log_2 P_Q$ when $L_A=-\log_2 P_Q$. We take  $L=400$, $p=0.08$ and $T_1=T_2=2L$.}
\end{figure}

The sublinear power-law exponent in the contiguous code distance is a special feature of the hybrid random dynamics and is closely related to the subleading correction term in the entanglement entropy. In the Clifford circuit, this can be easily understood in the dynamics of the stabilizer generators, in which there exist a finite number of ``short" stabilizers caused by local measurements\cite{PhysRevB.100.134306}. These short stabilizers are responsible for both the 
fluctuation in the entanglement entropy and also the sublinear power-law exponent in the code distance\cite{PhysRevB.103.104306}. Under pure unitary dynamics, these short stabilizers become long stabilizers and span over the whole system, the subleading correction term vanishes and the code distance becomes extensive and is proportional to $L$, the same as the conventional random QECC\cite{Brown_2013}. 

The above physics can also be understood in the hybrid QA circuit as shown in Fig. \ref{fig:UP_U_circuit}. Compared with Fig. \ref{fig:QA_puri}, we add an extra pure unitary evolution for time $T_2$.  Recall that the particle representation experiences the circuit dynamics in a time-reversed order, it first evolves under the pure unitary evolution for $T_2$ and then the hybrid dynamics for $T_1$. Here we take sufficiently long $T_2$ for unitary evolution so that  the particles are fully scrambled and only the configurations with no $X$ ($Y$) particles at the beginning can contribute to $P_X$ ($P_Y$). Hence, $P_X=2^{-L_A}$ and $P_Y=2^{L_A-L}$ and we have 
\begin{equation}
  S_A^{(2)}(t)=
  \begin{cases}
    L_A, & L_A< L^c \\
    L-L_A-\log_2 P_2(t), & L_A>L^c.
  \end{cases}
\end{equation}
Here $-\log_2 P_2(t)$ is simply counting the number of independent basis initially defined in $A$.

To verify this result, we simulate the Clifford QA circuit and compare the results with that derived from the particle model. As shown in Fig. \ref{fig:UP_U_comparison}, we find that $S_A^{(2)}$ agrees with $L_A$ for $L_A<L^c$ and $L-L_A-\log_2 P_2(t)$ for $L_A>L^c$ with negligible fluctuation. Different from the previous circuit defined in Fig.~\ref{fig:QA_puri}, it is easy to numerically evaluate $P_2(t)$ in this circuit. Due to the scrambling property of the unitary evolution in $T_2$, we find that over a large range of $L_A$, $-\log_2 P_2(t)=-\log_2 P_Q(t)$ and they become different only when $L_A<-\log_2 P_Q(t)$.
There is no subleading correction term in $S_A^{(2)}$ anymore and the code distance is $L-L^c$ which is linearly proportional to $L$. These results indicate that the sublinear power-law scaling in both the contiguous code distance and the fluctuation of the entanglement entropy are emergent properties of the hybrid random circuit and disappear when the dynamics is fully scrambled under unitary evolution.

\section{Classical linear code}\label{sec: classical code}


The classical particle model discussed in this paper has an interesting connection with the classical error correction.
For a system with $L$ sites, the total number of the particle configurations is $2^L$ and all of them can be generated from $L$ independent particle string basis.
Under the unitary dynamics, 
the number of basis is invariant, indicating that the total amount of the classical information is unchanged. On the other hand, the composite measurement forces $\bullet\to\circ$ at one site in all of the basis and can potentially reduce the number of independent basis, resulting in the loss of information. 

The information retained in the classical particle model can be characterized by the number of independent basis $k$. Under the purification dynamics in Sec. \ref{sec: QA_puri}, $k$ is the same as the entropy $S_Q$.  When $0<p<p_c$, it takes $\exp(L)$ time for $k(t)$ decreasing to zero. On the other hand, when $p>p_c$, $k(t)$ decreases to zero exponentially fast with a finite decay rate. The phase transition at $p=p_c$ belongs to the directed percolation universality class.  

\begin{figure}[tp!]
  \centering
  \subfigure[]{
    \includegraphics[width=0.45\textwidth]{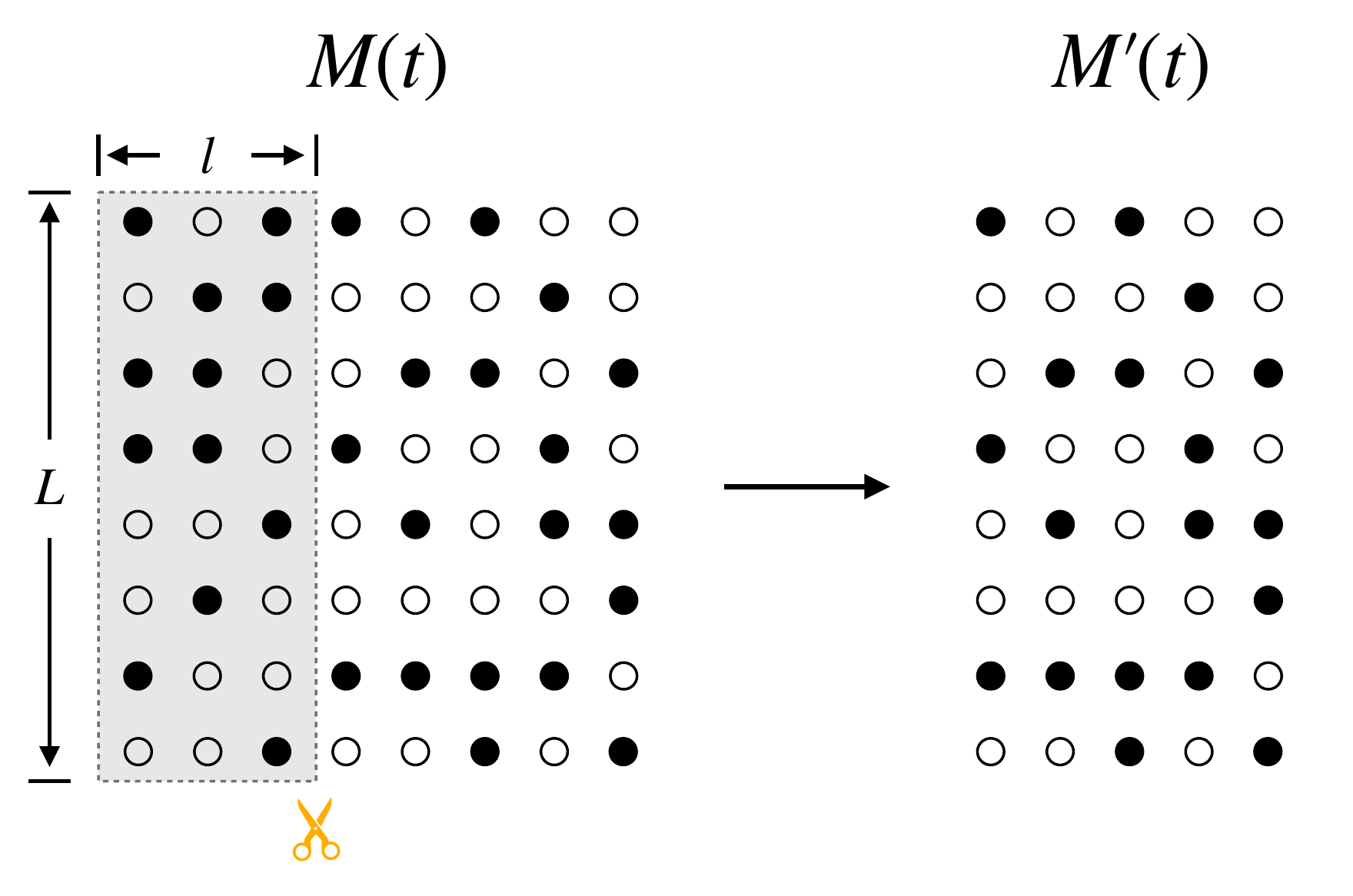}
    \label{fig:CLC_cartoon}
  }
  \subfigure[]{
    \includegraphics[width=0.4\textwidth]{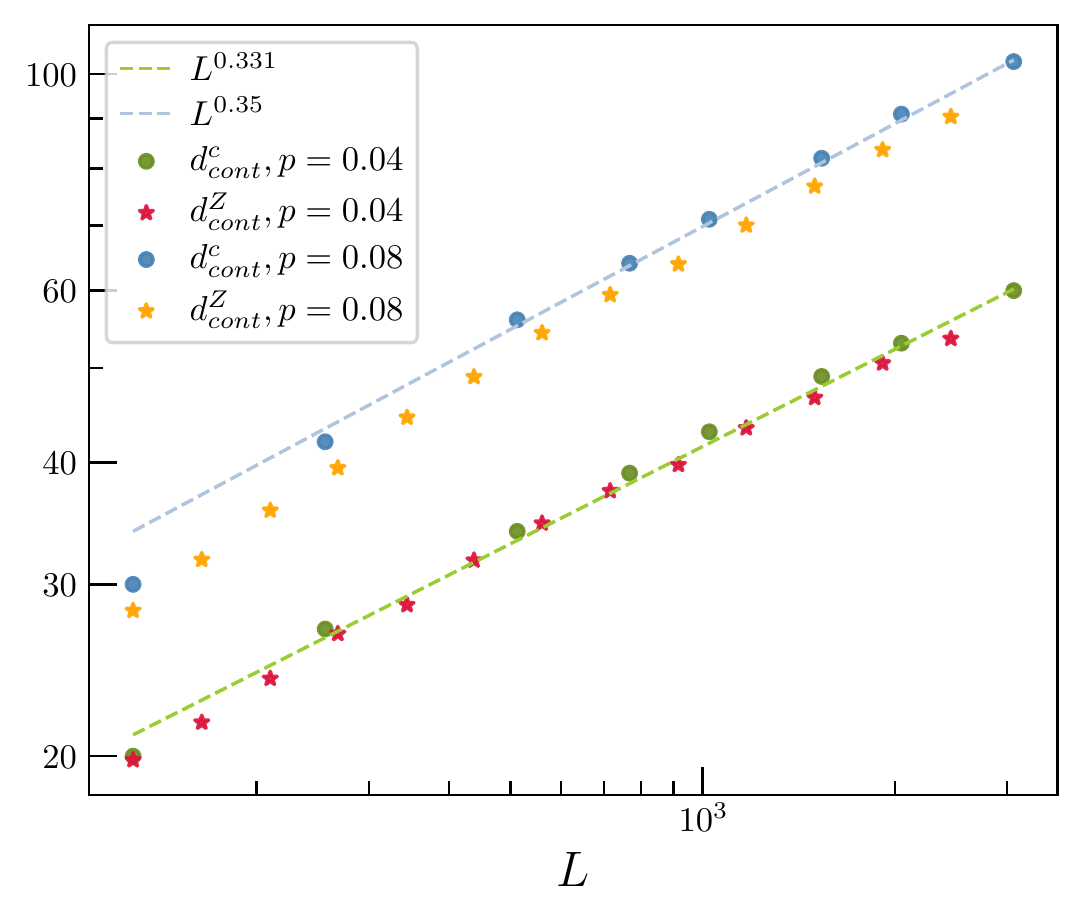}
    \label{fig:CLC_d}
  }
  
  \caption{(a) The CLC is determined by the binary square matrix $M(t)$ on the left. The occupied site symbol $\bullet$ denotes $1$ and the empty site symbol $\circ$ denotes $0$. To evaluate the contiguous CLC code distance $d^c_{\text{cont}}$ at time $t$, an $L\times l$ submatrix is taken away from $M(t)$ and $d^c_{\text{cont}}$ is the largest $l$ such that the rank of the truncated matrix $M'(t)$ agrees with that of $M(t)$ within $\epsilon$. In the numerical simulation, we set $\epsilon=1$. (b) $d^c_{\text{cont}}$ and $d^Z_{\text{cont}}$ vs system size $L$ for $p=0.04$ and $p=0.08$ at $T=4L$ plotted on a log-log scale.}
  
\end{figure}

Similar to the dynamically generated QECC with $p<p_c$, the associated particle dynamics also generates a classical linear code (CLC) governed by a $k\times L$ generator matrix, whose rows are binary strings forming a basis for the $k$ dimensional codespace. When the time $t$ is linear in $L$, the encoded bit $k$ is extensive and the information is protected by the scrambling property of the unitaries and is inaccessible by the local measurement. A CLC is typically denoted by $[L,k,d^c]$, in which $L$ classical bits can store $k$ bits of classical information. $d^c$ is the classical code distance and is equal to the minimal number of flips mapping a codeword to another. Similar to the QECC discussed before, since we have local unitary dynamics, it is more reasonable to consider contiguous code distance $d^c_{\text{cont}}$ for our CLC. 

In a CLC $[L,k,d^c_{\text{cont}}]$, any bit flip occurring in a subsystem with length $l<d^c_{\text{cont}}$ does not change the encoded bit $k$. Numerically, this motivates us to evaluate $d^c_{\text{cont}}$ in the following way as illustrated in Fig.~\ref{fig:CLC_cartoon}: consider an initial generator matrix $M$ with $\mbox{rank}_2 (M)=L$. We evolve all of the row vectors according to the hybrid QA circuit described in Fig.~\ref{fig:purification} (b). At any time $t$,  the encoded bit is the number of the independent binary vectors in $M(t)$, i.e., $k=\mbox{rank}_2 (M(t))$. We then remove a contiguous $L\times l$ submatrix from $M(t)$ and obtain a truncated $M^\prime(t)$. The largest $l$ which makes $k-\mbox{rank}_2 (M^\prime(t))<\epsilon$ is $d^c_{\text{cont}}$. In the numerical simulation, we take $\epsilon=1$ and we are interested in the regime $t=\alpha L$ with $\alpha\gg 1$. As shown in Fig.~\ref{fig:CLC_d}, we observe that $d^c_{\text{cont}}\propto L^{0.331}$ for $p=0.04$ and $d^c_{\text{cont}}\propto L^{0.35}$ for $p=0.08$. We also plot the $Z$-error QECC contiguous code distance $d^Z_{\text{cont}}$ for comparison. Although there is a slight difference in measuring these two quantities that the $Z$ error occurs at $t=0$ for the particle model due to its time-reversal evolving property, the numerics shows a resemblance between $d^Z_{\text{cont}}$ and $d^c_{\text{cont}}$. The diverging code distance is consistent with exponentially long purification time --- the information is encoded non-locally and is resilient to any local errors.

\section{Conclusion}
In this paper, we analyze entanglement entropy fluctuations in the volume-law phase of 1+1d hybrid QA circuits. We numerically show that the fluctuations belong to the KPZ universality class, just as for other random circuits studied previously. Due to the special feature of the QA circuit, we are able to map the second R\'enyi entropy to a classical quantity in a particle model. We compute the fluctuations of this quantity in different approaches and show that they exhibit fluctuations with similar exponents. The existence of the strong fluctuations may have interesting connection with the multifractal behavior observed in the volume-law phase of hybrid Clifford circuits, where the entanglement entropy transition is mapped to an Anderson localization transition\cite{Iaconis_2021}.
Besides this, we could also study fluctuations at the critical point in these hybrid random circuits. These critical points, dominated by randomness, are different from those clean systems. Since the hybrid QA circuit has an underlying particle picture, it could be a good starting point to explore this problem.

We also study the dynamically generated QECC in the purification dynamics of 1+1d hybrid QA circuits. Again, we give an interpretation of the error correction in terms of the particle model. In particular, we show that the particle model itself can be treated as a random classical linear code (CLC), and numerically compute the contiguous code distance for it. This observation motivates us to consider other random CLCs and use similar approaches to construct QECC. We leave this for the future study. 

\acknowledgements
We acknowledge Ethan Lake for his proofreading and acknowledge the useful discussion with Yaodong Li. We gratefully acknowledge computing resources from Research Services at Boston College and the assistance provided by Wei Qiu. This research
is supported in part by the Google Research Scholar Program.

\appendix
\section{Two-species particle model}\label{Appendix: 2ps}
In Ref.~\onlinecite{PhysRevB.105.064306}, we proposed a two-species BAW model which maps the entanglement dynamics of $\mathbb{Z}_2$-symmetric hybrid QA circuits to the classical dynamics of two kinds of particles performing branching-annihilating random walks (BAW). The two-species particle model can also be modified and applied on hybrid QA circuits without $\mathbb{Z}_2$ symmetry. Before introducing the particle model, we will first give an overview of the classical bit string dynamics.

Recall that the second R\'enyi entropy $S_A^{(2)}=-\log_2{\text{Tr}(\rho_A^2)}$. The purity $\text{Tr}(\rho_A^2)$ equals the expectation value of the $\mathsf{SWAP}_A$ operator over two copies of the state \cite{PhysRevLett.104.157201,islam2015measuring},
\begin{equation}
  \text{Tr}[\rho_A^2(t)]=\langle\psi_t|_2\otimes\langle\psi_t|_1 \mathsf{SWAP}_A|\psi_t\rangle_1\otimes|\psi_t\rangle_2.
\end{equation}
The wave function can be partitioned into subregions $A$ and $B$
\begin{equation}
  \begin{aligned}
    |\psi_t\rangle=\tilde{U}_t|\psi_0\rangle=\tilde{U}_t|+x\rangle^{\otimes L}=\frac{1}{\sqrt{2^L}}\sum_{i,j}e^{i\theta_{ij}}|\alpha_i\rangle_A|\beta_j\rangle_{B},
  \end{aligned}
\end{equation}
where $\tilde{U}_t=M_t U_t M_{t-1} U_{t-1}\cdots$ represents the hybrid QA circuit of depth $t$ as an alternating combination of layers of measurements and unitary evolution. The $\mathsf{SWAP}_A$ operator exchanges the spin configurations $|\alpha\rangle$ within subsystem $A$ of the double copies of $|\psi_t\rangle$. Then, we insert two sets of complete basis which we call ``bit strings'' \cite{Iaconis_2021pmf},
\begin{equation}\label{eq:purity}
  \begin{aligned}
    \text{Tr}[\rho_A^2(t)]&=\sum_{n_1,n_2}\langle\psi_t|_2\langle\psi_t|_1 \mathsf{SWAP}_A|n_1\rangle|n_2\rangle\langle n_2|\langle n_1|\psi_t\rangle_1|\psi_t\rangle_2
    \\
    &=\sum_{n_1,n_2}\langle\psi_0|_1\tilde{U}_t^{\dagger}|n_1'\rangle\langle\psi_0|_2\tilde{U}_t^{\dagger}|n_2'\rangle
    \\
    & \qquad \qquad \langle n_1|\tilde{U}_t|\psi_0\rangle_1\langle n_2|\tilde{U}_t|\psi_0\rangle_2
    \\
    &=\frac{1}{4^L}\sum_{n_1,n_2}e^{-i\Theta_{n_1'}(t)}e^{-i\Theta_{n_2'}(t)}e^{i\Theta_{n_1}(t)}e^{i\Theta_{n_2}(t)},
  \end{aligned}
\end{equation}
where 
\begin{equation}
  \begin{aligned}
    |n_1'\rangle|n_2'\rangle&=\mathsf{SWAP}_A|n_1\rangle|n_2\rangle
    \\
    &=\mathsf{SWAP}_A|\alpha_1\beta_1\rangle|\alpha_2\beta_2\rangle
    \\
    &=|\alpha_2\beta_1\rangle|\alpha_1\beta_2\rangle.
  \end{aligned}
\end{equation}
Strictly speaking, there does not exist $\tilde{U}_t^\dagger$ since the projective measurements are nonunitary operators. However, we can still deduce the effective action of the composite measurement on the bit string,
\begin{equation}
  \begin{aligned}
    \langle n|& M_i^{\sigma}|\psi_0\rangle =\langle n| H\circ P_i^\sigma|\psi_0\rangle
    \\
    &=\langle T_i^{\sigma}(n)|\psi_0\rangle = \frac{1}{\sqrt{2^L}}e^{i\theta_{T_i^{\sigma}(n)}},
  \end{aligned}
\end{equation}
where $\langle T_i^\sigma (n)|$ stands for the bit string $\langle n|$ with the spin on site $i$ forced to be in the $\sigma$ state. Hence, instead of following the quantum trajectory of $|\psi_t\rangle$, we can study the bit string dynamics in a time-reversed order, i.e., evaluate $\langle n|\tilde{U}_t|\psi_0\rangle$ from left to right,
\begin{equation}
  \begin{aligned}
    \langle n|\tilde{U}_t|\psi_0\rangle &=\langle n(t'=0)|M_t U_t M_{t-1} U_{t-1}\cdots|\psi_0\rangle
    \\
    &=\langle n(t'=1)|U_t M_{t-1} U_{t-1}\cdots|\psi_0\rangle
    \\
    &=e^{i\theta_{n(t'=1)}}\langle n(t'=1)|M_{t-1} U_{t-1}\cdots|\psi_0\rangle
    \\
    &\cdots
    \\
    &=\frac{1}{\sqrt{2^L}}e^{i\theta_{n(t'=1)}}e^{i\theta_{n(t'=2)}}\cdots e^{i\theta_{n(t'=t)}}
    \\
    &=\frac{1}{\sqrt{2^L}}e^{i\Theta_n(t)},
  \end{aligned}
\end{equation}
where $e^{i\Theta_n(t)}$ is one of the accumulated phase terms under time evolution that are multiplied and summed up over the ensemble of all the possible bit-string configurations $\{|n_1\rangle,|n_2\rangle,|n_1'\rangle,|n_2'\rangle\}$ in Eq. \ref{eq:purity} to evaluate $\text{Tr}\rho_A^2$.

\begin{figure}
  \centering
  \includegraphics[width=0.45\textwidth]{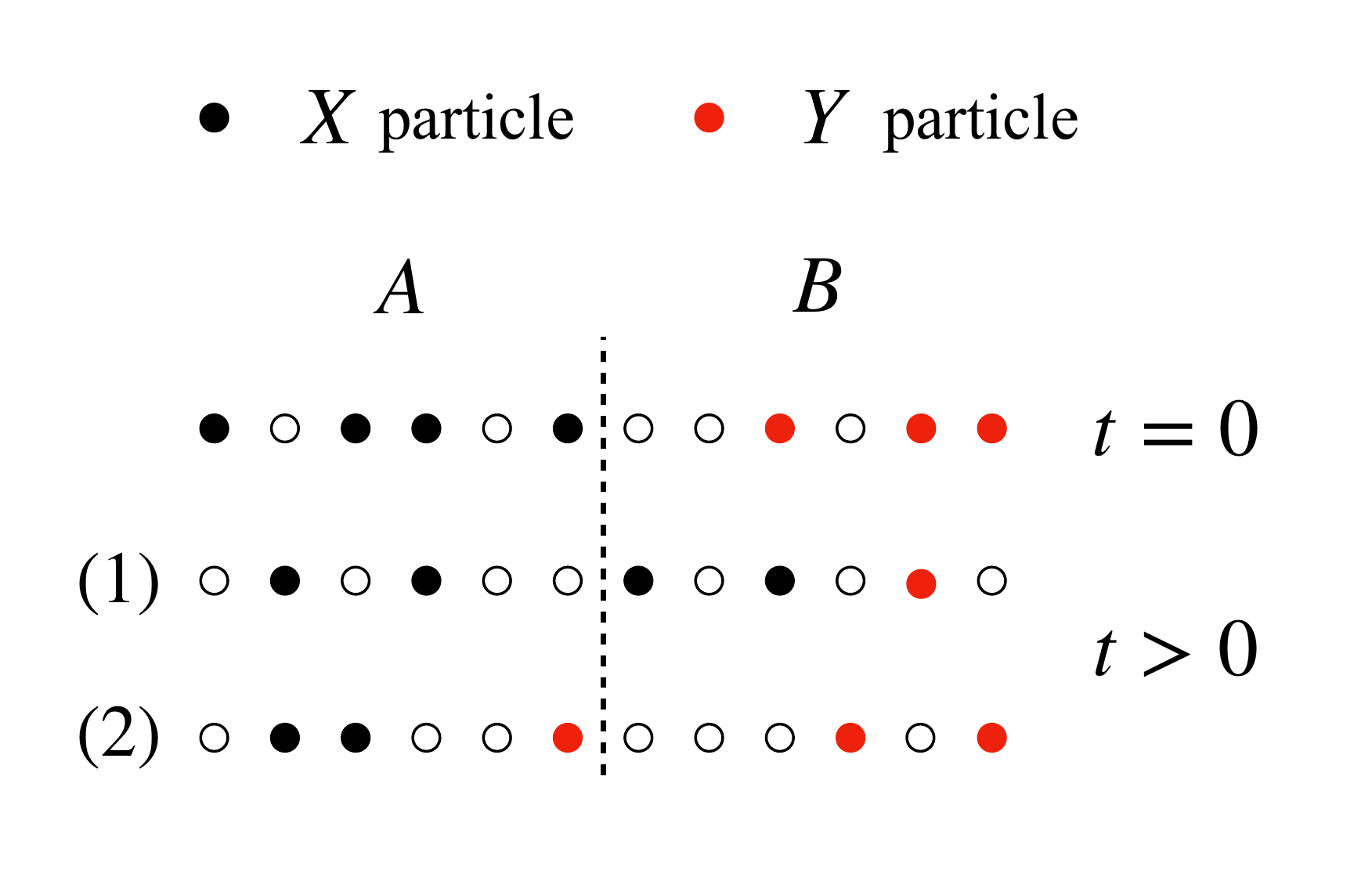}
  \caption{An example of the two-species particle model. The black dots represent $X$ particles, and the red dots represent $Y$ particles. Initially, $X$ and $Y$ particles are distributed in region $A$ and $B$ respectively. Under the time evolution, the two species expand according to the single-species update rule before they encounter one another. There are two types of possible particle configurations in which the two species have not met up to time $t$: (1) $X$ particles intrude into $B$ and (2) $Y$ particles intrude into $A$.}
  \label{fig:2ps_cartoon}
\end{figure}

In order to understand the dynamics of the relative phase $\Theta_r=-\Theta_{n_1'}-\Theta_{n_2'}+\Theta_{n_1}+\Theta_{n_2}$, we consider the evolution of the difference between a bit string pair $\{|n_1\rangle,|n_2\rangle\}$,
\begin{equation}
  h(x,t)=|n_1(x,t)-n_2(x,t)|.
\end{equation}
It is then natural to use the particle representation where the empty site symbol $\circ$ denotes $h(x)=0$ and the occupied site symbol $\bullet$ denotes $h(x)=1$. Specifically, we represent the difference at $t=0$ in subregion $A$ (subregion $B$) by $X$ ($Y$) particles. It is easy to check that within the regime occupied by particles of the same species, under CNOT gate with the first qubit acting as the control, $\bullet\circ\leftrightarrow\bullet\bullet$, i.e., the particle gives birth to another particle of the same kind at the neighboring site or kills another if the neighbor is already occupied. On the other hand, under the composite measurement, $\bullet\to\circ$, i.e., the particle annihilates with probability $p$. Let $x$ denote the position of the rightmost $X$ particle and $y$ denote the position of the leftmost $Y$ particle. As shown in Fig. \ref{fig:2ps_cartoon}, under the time evolution, the particles start to evolve according to the update rule. Meanwhile, $x$ and $y$ can also be viewed as the ``end particles'' performing biased random walks. Before the two species encounter each other, the phase generated by each layer of unitary evolution on $|n\rangle$ is $\theta_n=\theta_n^{[1,x]}+\theta_n^{(x,y)}+\theta_n^{[y,L]}$, i.e., the sum of phases generated within the regimes $[1,x]$, $(x,y)$ and $[y,L]$. The bit string configurations within $[1,x]$ occupied by $X$ particles always satisfy $n_1([1,x])=n_2'([1,x])$ and $n_2([1,x])=n_1'([1,x])$. Therefore, $\theta_{n_1}^{[1,x]}=\theta_{n_2^\prime}^{[1,x]}$ and $\theta_{n_2}^{[1,x]}=\theta_{n_1^\prime}^{[1,x]}$. Similarly, for the regime occupied by $Y$ particles, since $n_1([y,L])=n_1'([y,L])$ and $n_2([y,L])=n_2'([y,L])$, we always have $\theta_{n_1}^{[y,L]}=\theta_{n_1^\prime}^{[y,L]}$ and $\theta_{n_2}^{[y,L]}=\theta_{n_2^\prime}^{[y,L]}$. At the same time, since there is no bit string difference within the regime $(x,y)$, $\theta_{n_1}^{(x,y)}=\theta_{n_2}^{(x,y)}=\theta_{n_1^\prime}^{(x,y)}=\theta_{n_2^\prime}^{(x,y)}$. Therefore, the phase difference along the lattice vanishes: $-\theta_{n_1'}-\theta_{n_2'}+\theta_{n_1}+\theta_{n_2}=0$. If for a bit string pair $\{|n_1\rangle,|n_2\rangle\}$, $X$ and $Y$ particles do not meet each other up to time $t$, then the accumulated relative phase $\Theta_{r}(t)$ is zero and such pair contributes $1/4^L$ to the purity $\text{Tr}[\rho_A^2(t)]$.

Once the rightmost $X$ particle comes across the leftmost $Y$ particle, the two-qubit phase gate acting on sites $x$ and $y$ will generate a nonzero relative phase. For example, if we apply the CZ gate on $\bullet\red{\bullet}$ with a possible corresponding bit string configuration $\{|n_1\rangle,|n_2\rangle,|n_1'\rangle,|n_2'\rangle\}_{x,y}=\{|10\rangle,|01\rangle,|00\rangle,|11\rangle\}$, a relative phase $0+0-0-\pi=-\pi$ is generated. If we apply the CNOT gate on sites $x$ and $y$, $\{|n_1\rangle,|n_2\rangle,|n_1'\rangle,|n_2'\rangle\}_{x,y}\to\{|11\rangle,|01\rangle,|00\rangle,|10\rangle\}$, i.e., another type of ``particle'' different from the two species with bit string configuration $|n_1\rangle_{y}=|n_2\rangle_{y}\neq|n_1'\rangle_{y}=|n_2'\rangle_{y}$ appears on site $y$ and will spread along the lattice under further evolution. As time evolves, the configurations for which the two species have met will generate random accumulated phases, half of which are composed of odd numbers of $\pi$, while the other half are composed of even numbers of $\pi$. The accumulated phase terms $e^{i\Theta_r}$ of such configurations will add up to zero and make no contribution to Eq. \ref{eq:purity}. Therefore, we have 
\begin{equation}
  \begin{aligned}
    &\text{Tr}\rho_A^2(t)\approx P(t),
    \\
    &S_A^{(2)}(t)\approx -\log_2{P(t)},
  \end{aligned}
  \end{equation}
where $P(t)$ is the fraction of particle configurations in which $X$ and $Y$ particles never encounter one another up to time $t$. This quantum-classical correspondence has been numerically verified in Ref.~\onlinecite{PhysRevB.105.064306}.

\section{Entanglement dynamics in the volume-law phase of $\mathbb{Z}_2$ symmetric hybrid Clifford QA circuit}\label{Appendix: Z_2}
  \begin{figure}
  \centering
  \subfigure[]{
    \includegraphics[width=.3\textwidth]{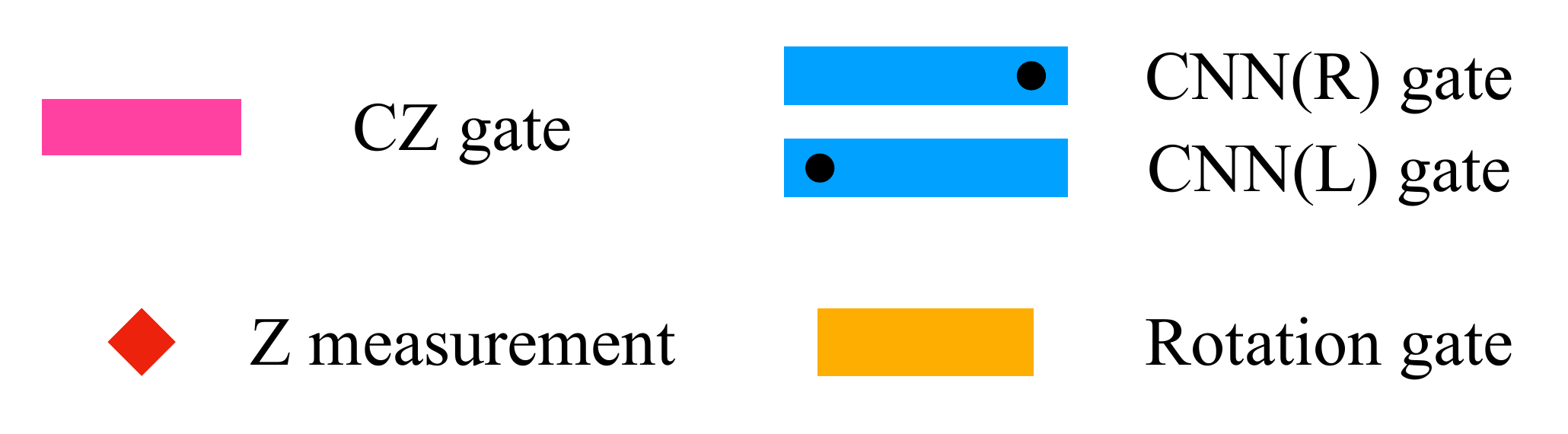}
    \label{fig:EE_gate_Z2}}
  \subfigure[]{
    \includegraphics[width=.25\textwidth]{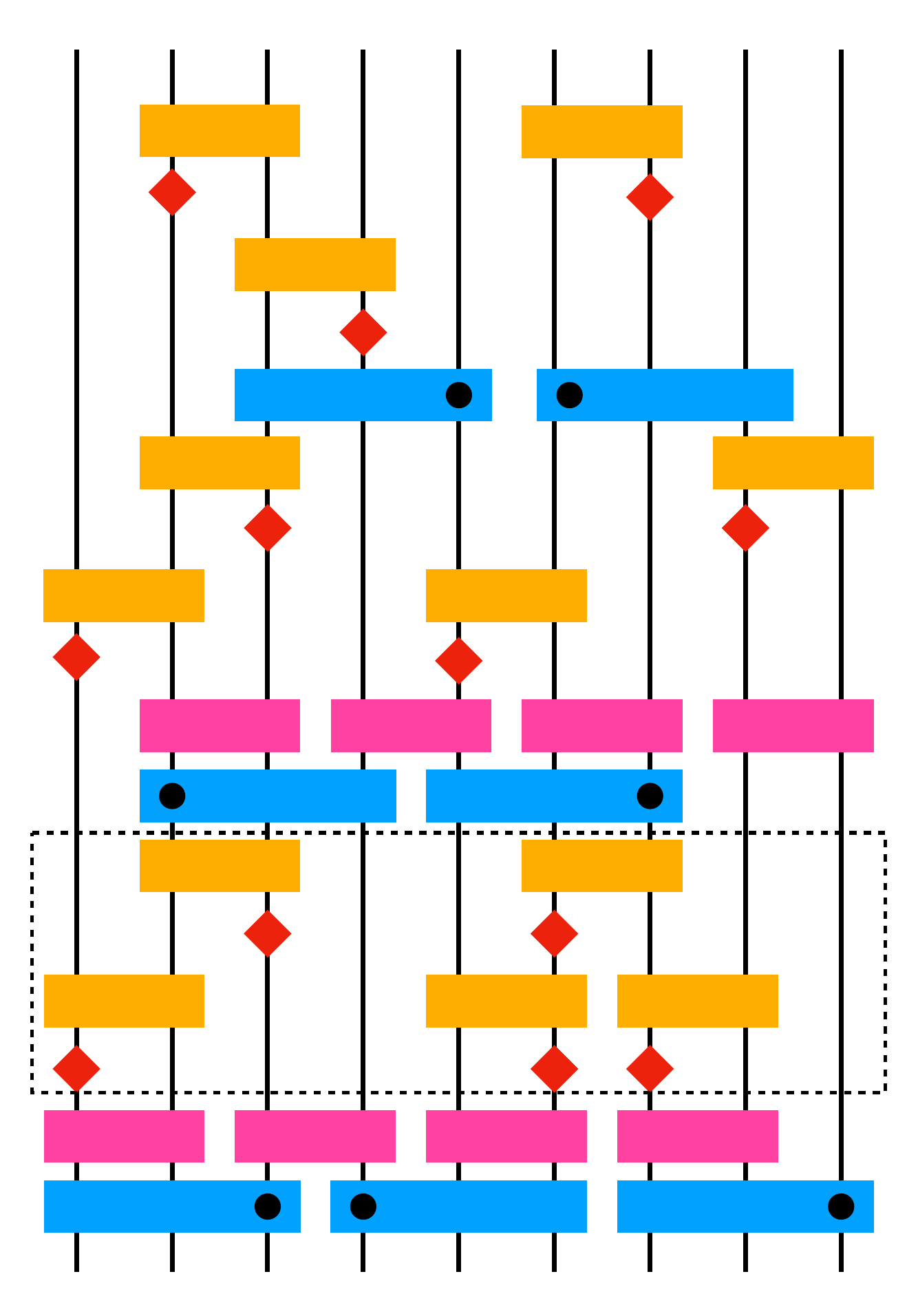}
    \label{fig:QA_EE_Z2}}

  \caption{(a) A schematic for the gates appearing in the circuit.  (b)  The arrangement of gates in a single time step of the $\mathbb{Z}_2$-symmetric hybrid QA circuit. Each time step involves three layers of CNN gates and two layers of CZ gates, interspersed with three measured layers. The dashed box represents a measured layer enclosing two rows of composite measurements, with the first/second row containing randomly distributed $M_{L/R}^{\sigma}$ applied on sites $(2i-1,2i)$[$(2i,2i+1)$] for $i\in[1,L/2]$. As with the CNN gates, the projection of $M_{L/R}^{\sigma}$ is chosen to be applied on the left/right qubit with equal probability. In general, the composite measurement appears in a measured layer with probability $p$.}
  \label{fig:entanglement_Z2}
  \end{figure}
  
In this appendix, we study the subleading correction term of the volume-law phase entanglement entropy of the $\mathbb{Z}_2$-symmetric hybrid QA circuit. The $\mathbb{Z}_2$ symmetry requires that the parity of the computational basis remains fixed. This can be satisfied by measuring the Pauli string $Z_1Z_2\dots Z_L$ on an initial product state with $L$ qubits polarized in the $+x$ direction. We choose a subset of Clifford gates to construct the QA circuit with $\mathbb{Z}_2$ symmetry and the setup is shown in Fig. \ref{fig:entanglement_Z2}. The unitary evolution composed of CNOTNOT(CNN) gates and CZ gates. The CNN gate flips two qubits according to the value of the third (control) qubit. If the control qubit is on the left we denote the corresponding gate as CNN$_L$; it acts as 
\begin{equation}
\begin{aligned}
  &\text{CNN}_L|1\sigma_1\sigma_2\rangle =|1(1-\sigma_1)(1-\sigma_2)\rangle
  \\
  &\text{CNN}_L|0\sigma_1\sigma_2\rangle=|0\sigma_1\sigma_2\rangle.
\end{aligned}
\end{equation}
Aside from the unitary evolution, we also introduce into the circuit the two-qubit composite measurements defined as 
\begin{equation}
M_{L/R}^{\sigma}=R\circ P_{L/R}^{\sigma}.
\end{equation}
This measurement is a combination of the projection operator  $P_{L/R}^{\sigma}$ on the left/right qubit into the spin $\sigma=\{0,1\}$, together with a two-site rotation operation
\begin{equation}
  R=\frac{1}{\sqrt{2}}\begin{pmatrix}
  1 & 0 & 0 & 1 \\
  0 & 1 & 1 & 0 \\
  0 & 1 & -1 & 0 \\
  1 & 0 & 0 & -1
\end{pmatrix},
\end{equation} 
so that the wave function is always an equal weight superposition of $\mathbb{Z}_2$ symmetric computational basis.

\begin{figure}[tp!]
  \centering
  \subfigure[]{
    \includegraphics[width=0.4\textwidth]{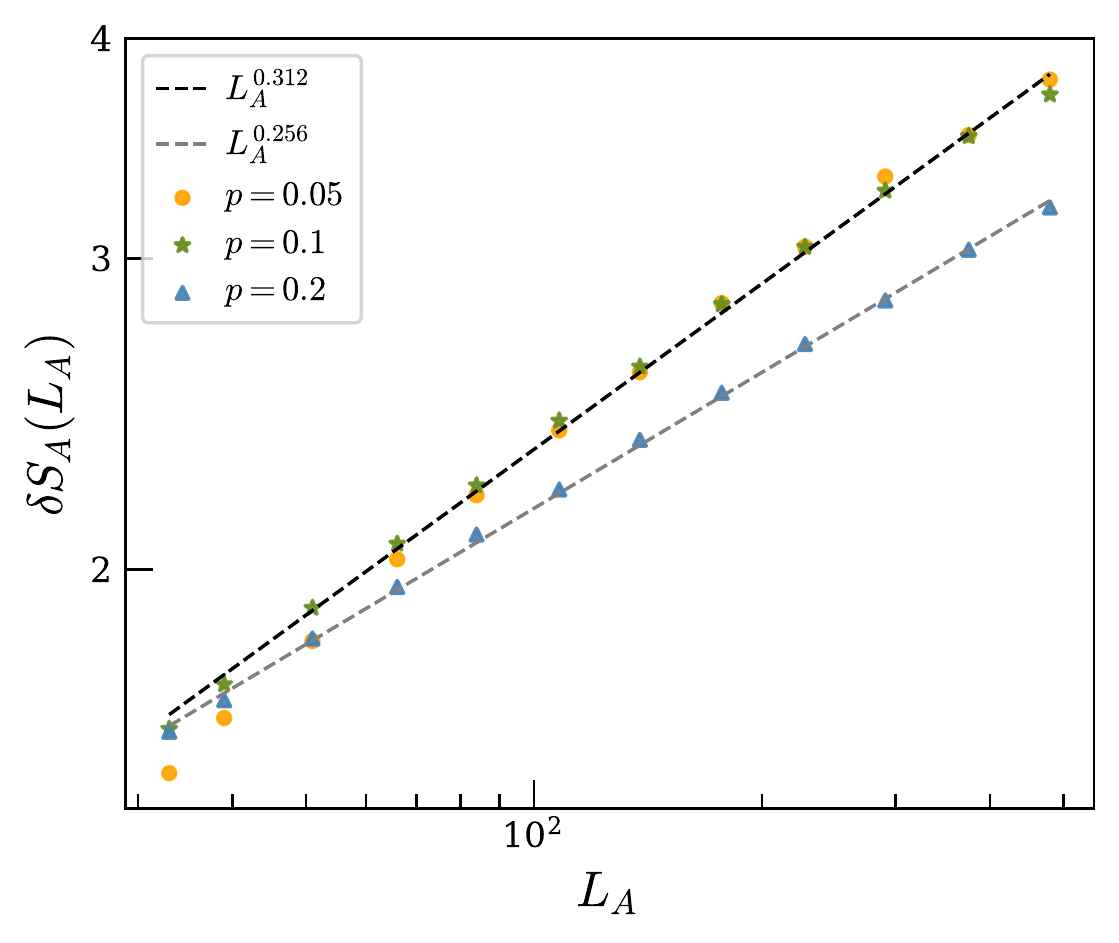}
    \label{fig:EE_Z2_LA}}
  \subfigure[]{
    \includegraphics[width=0.4\textwidth]{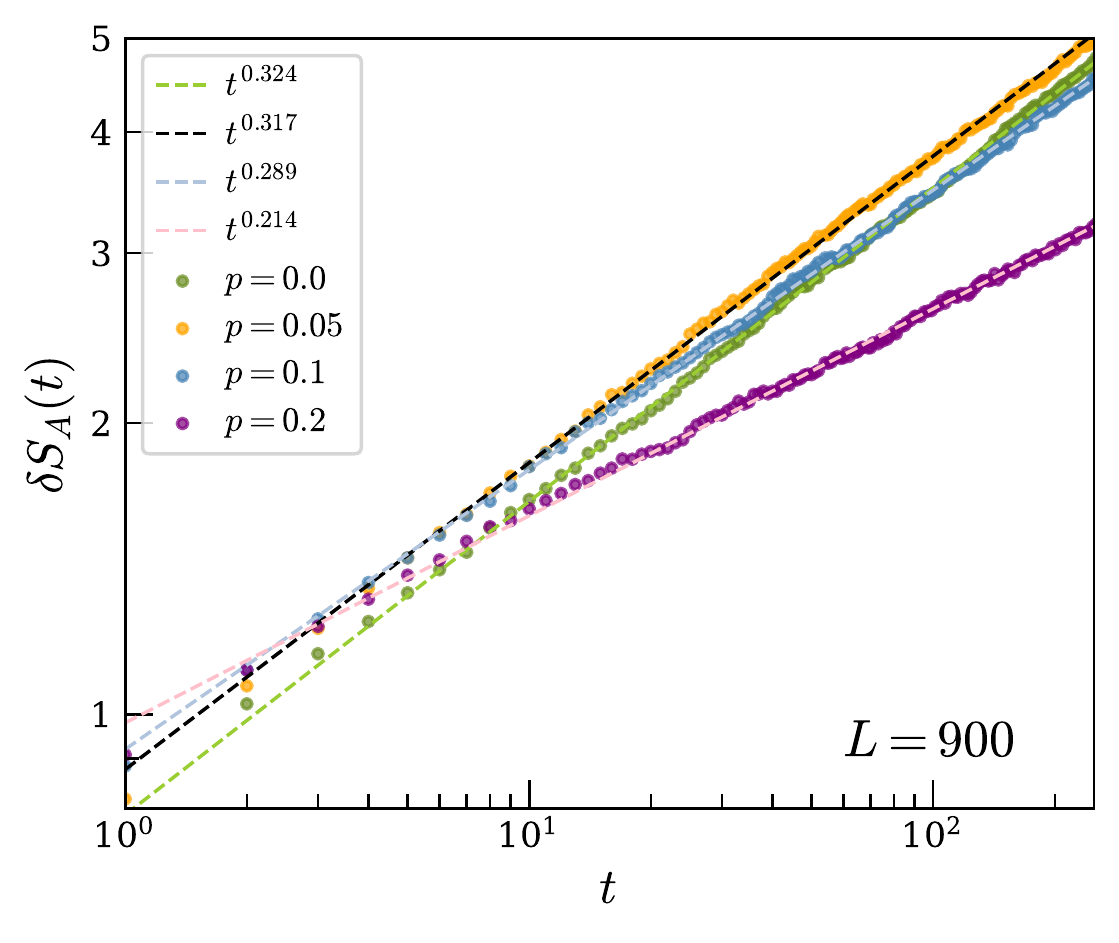}
    \label{fig:EE_Z2_t}}
  \caption{The standard deviation of entanglement entropy of the $\mathbb{Z}_2$-symmetric hybrid Clifford QA circuit. (a) $\delta S_A$ vs $L_A$ plotted on a log-log scale. The data are computed from the steady-state entanglement entropy $S_A$ for half-system size $L_A=L/2$ over a variety of $L$. The measurement rates are taken to be $p=0.05,0.1,0.2$. (b) $\delta S_A$ vs $t$ for $p=0,0.05,0.1,0.2$. All of the numerical data for entanglement entropy are calculated with periodic boundary conditions (PBC).}
\end{figure}

As shown in Ref.~\onlinecite{PhysRevB.105.064306}, the competition of the unitary evolution and the composite measurements leads to an entanglement phase transition from a volume-law phase to a critical phase as the measurement rate $p$ increases, and the critical point is $p_c=0.335$. Here we focus on the subleading correction of the entanglement entropy in the volume-law phase $p<p_c$. We first compute the fluctuation of the steady state entanglement entropy. As shown in Fig. \ref{fig:EE_Z2_LA}, $\delta S_A\propto L_A^{\beta_1}$ with $\beta_1=0.312$ for $p=0.05$ and $p=0.1$, $\beta_1=0.256$ for $p=0.2$. In Fig. \ref{fig:EE_Z2_t}, we compute the fluctuation of the early time entanglement entropy and find that $\delta S_A\propto t^{\beta_2}$ with $\beta_2=0.324$ for $p=0$, $\beta_2=0.317$ for $p=0.05$, $\beta_2=0.289$ for $p=0.1$ and $\beta_2=0.214$ for $p=0.2$. Similar to the case in the QA circuit without $\mathbb{Z}_2$ symmetry, the fluctuation exponents exhibit a drop from the roughness exponent $\beta=\frac{1}{3}$ as $p$ approaches $p_c$.

\section{Single-species end-point RWRE model}\label{Appendix:RWRE}

TD-RWRE refers to random walks performed in a $\emph{fixed}$ random environment. Different from the diffusions in $\emph{time-independent}$ random media where the fluctuations are of order $\sqrt{t}$, it was found that in the large deviations regime of TD-RWRE, fluctuations of the logarithm of the transition probability are distributed with the growth exponent $\beta=\frac{1}{3}$ of the DPRE, i.e.
\begin{equation}
  \log_2{P(X_t>ut)}\sim C_1(u)t+C_2(u)t^{\frac{1}{3}}\chi
\end{equation}
at large time, where $u>u_c=0$ and $\chi$ obeys the GUE Tracy-Widom distribution \cite{Corwin_2017wa,Barraquand_2017vr,PhysRevE.96.010102}. Hence, the large deviations regime of TD-RWRE belongs to the KPZ universality class.

In the two-species particle model, the rightmost $X$ particle and the leftmost $Y$ particle can be regarded as two end-point particles performing TD-RWRE since all the configurations experience the same circuit dynamics. To unravel the problem, we consider the single-species particle model introduced in Sec.\ref{sec: 1ps}, in which we focus on the phase difference of $|n_1\rangle$ and $|n_1'\rangle$ in the $B$ region measured by the quantity
\begin{equation}
  \frac{1}{4^{L_A}}\sum_{\alpha_1,\alpha_2}e^{-i\Theta_{n_1'}^{B}}e^{i\Theta_{n_1}^{B}}.
\end{equation}
Denoting the bit-string difference $|n_1-n_1'|$ as particles, it is obvious that this quantity equals $K(t)$ which is the fraction of configurations in which the particles initially located in regime $A$ never cross the boundary between $A$ and $B$ up to time $t$. Therefore, we only care about the dynamics of the end points of each particle configuration and can treat them as biased random walkers in a fixed random environment.

\begin{figure}[tp!]
  \centering
  \includegraphics[width=0.4\textwidth]{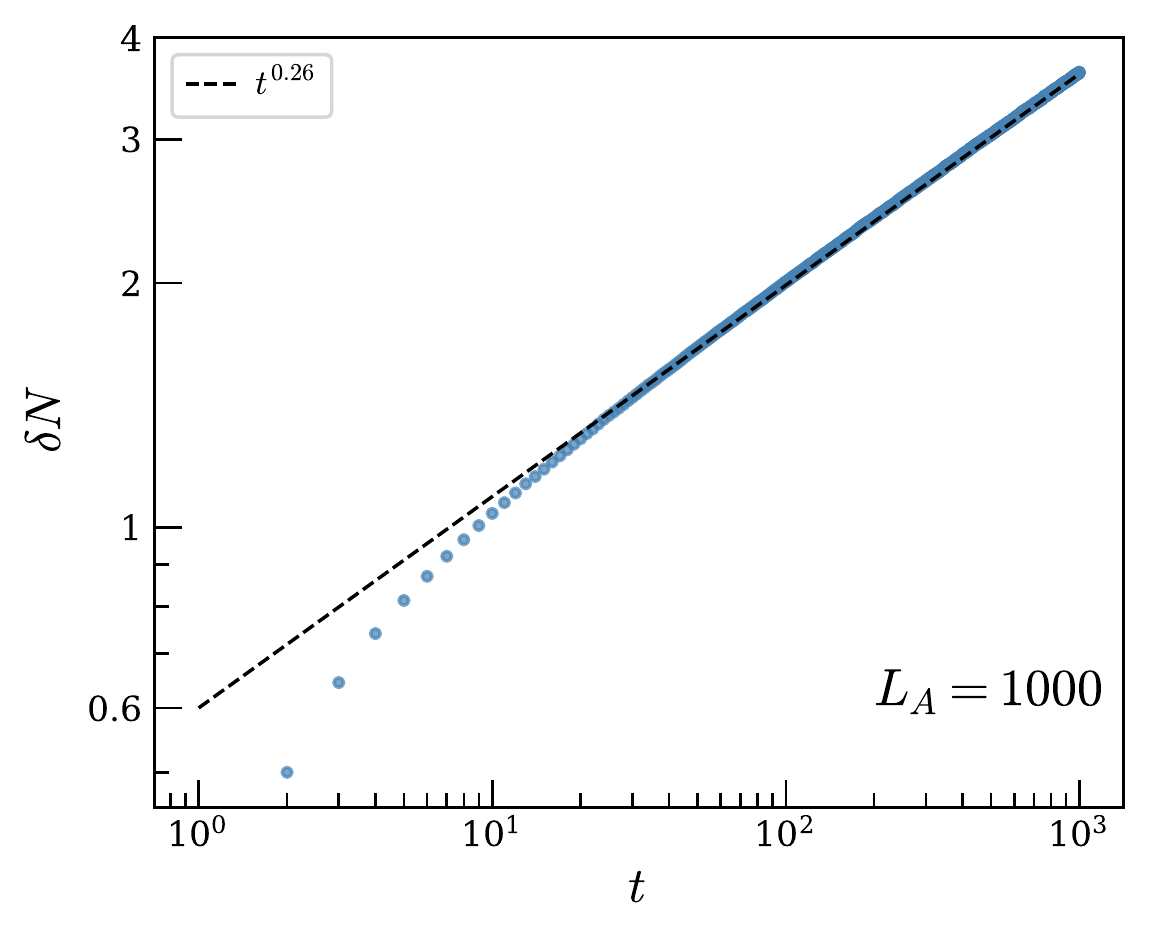}
  \caption{The standard deviation $\delta N$ vs $t$ for $L_A=1000$. }
  \label{fig:RWRE}
\end{figure}

Based on the above analysis, we propose a single-species end-point RWRE model. Initially, we place the end points of all the particle configurations on the lattice, which results in a lattice chain fully occupied in $A$ and empty in $B$. We further simplify the problem by assuming that the configurations with the same end point $\emph{initially}$ share the same dynamics, so that each site can be viewed as being occupied by only one particle at $t=0$. At each time step, a random value $\omega_i\in(0,1)$ is assigned to each site $i$ on which the particles have the probability $\omega_i$ to move to the right. Assume that when an end point originally located on site $i$ arrives at the boundary, the end points originally sit on the right of $i$ have already arrived. Define $N(t)$ as the number of particles that have already passed the boundary at time $t$, the ``entanglement entropy'' can be expressed as
\begin{equation}
    \begin{aligned}
      -\log_2{K(t)}&\approx -\log_2{\frac{2^{L_A}-2^{L_A-1}-\cdots 2^{L_A-N(t)}}{2^{L_A}}}\\
      &= -\log_2{\frac{2^{L_A-N(t)}}{2^{L_A}}}=N(t).
    \end{aligned}
\end{equation}

$N(t)$ grows linearly in time and eventually saturates to $L_A$. In Fig. \ref{fig:RWRE}, we compute the standard deviation $\delta N(t)$ and find that it scales as $t^{0.26}$.

\section{Purification dynamics in the volume-law phase}\label{Appendix: Clifford puri}
\begin{figure}[htp!]
  \centering
  \subfigure[]{
    \includegraphics[width=0.4\textwidth]{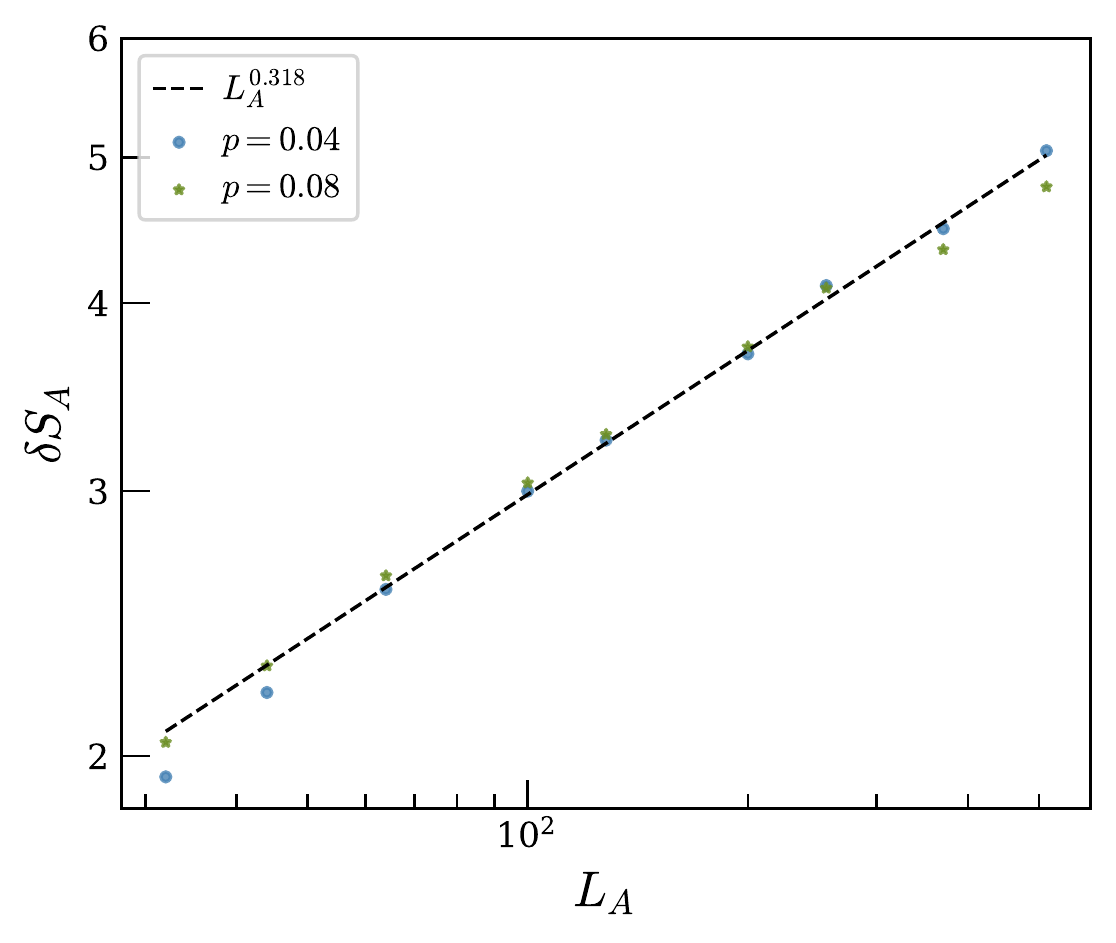}
    \label{fig:puri_dS}
  }
  \subfigure[]{
    \includegraphics[width=0.4\textwidth]{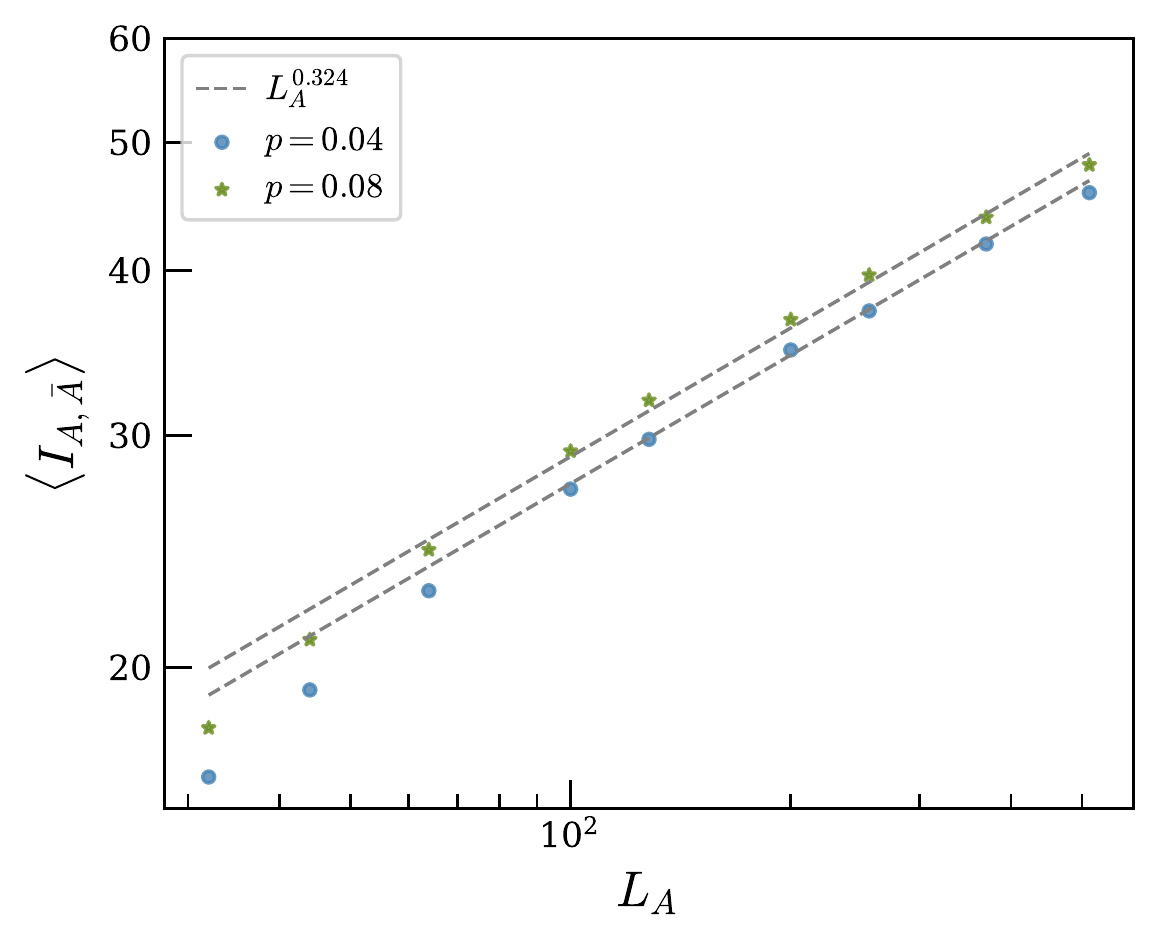}
    \label{fig:puri_I_AB}
  }
  \caption{(a) The standard deviation of the entanglement entropy $\delta S_A$ vs $L_A$ plotted on a log-log scale. (b) The mutual information between the two subsystems $I_{A,B}$ vs $L_A$ plotted on a log-log scale. All of the data are collected from the hybrid Clifford QA circuit at half system size $L_A=L/2$ for a variety of $L$ at $T=3L$ and are computed under PBC.}
\end{figure}

In this appendix, we numerically study the fluctuation exponent of the purification process of the hybrid Clifford QA model in the volume-law phase $p<p_c$. We first compute the fluctuation of the entanglement entropy of subsystem $A$ in Fig. \ref{fig:puri_dS} and find that $\delta S_A\propto L_A^{0.318}$, with the exponent $0.318$ close to the roughness exponent. We can also extract the subleading term out by computing the mutual information between the two subsystems $I_{A,B}=S_A+S_{B}-S_Q$. As shown in Fig. \ref{fig:puri_I_AB}, $I_{A,B}\propto L_A^{0.324}$. This again indicates the correlation between the volume-law phase of hybrid QA circuits and the KPZ universality class.

\section{Two-species particle model of the purification process}\label{Appendix:2ps_puri}
In order to interpret the purification process in terms of the two-species particle model, we go back to the bit string picture and modify Eq. \ref{eq:purity}. The wave function can now be expanded in the basis in subsystems $A$ and $B$ and also the environment $R$,
\begin{equation}
  \begin{aligned}
    |\psi_t\rangle&=\tilde{U}_t\circ \text{CZ}|\psi_0\rangle
    \\
    &=\tilde{U}_t\circ \text{CZ}|+x\rangle^{\otimes 2L}
    \\
    &=\frac{1}{\sqrt{4^{L}}}\sum_{i,j,k}e^{i\theta_{i,j,k}}|\alpha_i\rangle_A|\beta_j\rangle_{B}|\gamma_k\rangle_R,
  \end{aligned}
\end{equation}
where the CZ gate acts on both the system and the environment, creating $L$ EPR pairs, and the following $\tilde{U}_t$ is the combination of the hybrid QA circuit applied solely on system $Q$. To compute the purity, we can still apply the $\mathsf{SWAP}_A$ operator which exchanges the spin configurations $|\alpha\rangle$ within subsystem $A$ of the replicated states, and insert two complete sets of basis upon which the operators act in a time-reversed order,
\begin{equation}\label{eq:puri_purity}
  \begin{aligned}
    \text{Tr}(\rho_A^2)&=\sum_{n_1,n_2}\langle\psi_t|_2\langle\psi_t|_1 \mathsf{SWAP}_A|n_1\rangle|n_2\rangle\langle n_2|\langle n_1|\psi_t\rangle_1|\psi_t\rangle_2
    \\
    &=\sum_{n_1,n_2}\langle\psi_0|_1 \text{CZ}\circ\tilde{U}_t^{\dagger}|n_1'\rangle\langle\psi_0|_2\text{CZ}\circ\tilde{U}_t^{\dagger}|n_2'\rangle
    \\
    & \qquad \qquad \langle n_1|\tilde{U}_t\circ\text{CZ}|\psi_0\rangle_1\langle n_2|\tilde{U}_t\circ\text{CZ}|\psi_0\rangle_2
    \\
    &=\frac{1}{4^{2L}}\sum_{n_1,n_2}e^{-i(\Delta_{n_1'}+\Theta_{n_1'})}e^{-i(\Delta_{n_2'}+\Theta_{n_2'})}
    \\
    & \qquad \qquad \qquad \times e^{i(\Delta_{n_1}+\Theta_{n_1})}e^{i(\Delta_{n_2}+\Theta_{n_2})},
  \end{aligned}
\end{equation}
where 
\begin{equation}
  \begin{aligned}
    |n_1'\rangle|n_2'\rangle&=\mathsf{SWAP}_A|n_1\rangle|n_2\rangle
    \\
    &=\mathsf{SWAP}_A|\alpha_1\beta_1\gamma_1\rangle|\alpha_2\beta_2\gamma_2\rangle
    \\
    &=|\alpha_2\beta_1\gamma_1\rangle|\alpha_1\beta_2\gamma_2\rangle.
  \end{aligned}
\end{equation}
Here $\Theta_n$ is the accumulated phase generated by the circuit within system $Q$ of the bit string $|n\rangle$, and $\Delta_n$ is the phase generated by the CZ gate acting on both $Q$ and $R$ of the time-evolved bit string $\tilde{U}_t|n\rangle$. 

Based on the analysis in Appendix \ref{Appendix: 2ps}, only the bit string configurations $\{|n_1\rangle,|n_2\rangle,|n_1'\rangle,|n_2'\rangle\}$ whose total accumulated phases are zero can contribute to $\text{Tr}(\rho_A^2)$. We can take a further step by assuming that only the configurations satisfying $\Delta_r=-\Delta_{n_1'}-\Delta_{n_2'}+\Delta_{n_1}+\Delta_{n_2}=0$ and $\Theta_r=-\Theta_{n_1'}-\Theta_{n_2'}+\Theta_{n_1}+\Theta_{n_2}=0$ contribute to the purity. The former constraint is met when $|n_1(t)\rangle=|\alpha_1\beta_1\gamma_1\rangle=|n_1'(t)\rangle=|\alpha_2'\beta_1'\gamma_1\rangle$, and $|n_2(t)\rangle=|\alpha_2\beta_2\gamma_2\rangle=|n_2'(t)\rangle=|\alpha_1'\beta_2'\gamma_2\rangle$. In the particle language, it means that the particles representing the bit-string difference $|n_1-n_1'|$ completely die out at time $t$. Meanwhile, the latter constraint is the same as in the entanglement dynamics, i.e., the $X$ and $Y$ particles representing the difference $|n_1(x,0)-n_2(x,0)|$ in $A$ and $B$ respectively never encounter each other up to time $t$. To summarize, we only need to count the configurations for which $X$ and $Y$ particles do not meet and $X$ particles have become extinct at time $t$. Let the fraction of such configurations be $P_1$, the entanglement entropy of the subsystem $A$ is then
\begin{equation}
  S_A^{(2)}(t)\approx -\log_2{P_1(t)}.
\end{equation}


\end{document}